\documentclass[twocolumn,times]{aastex62}

\usepackage{textcomp}
\usepackage{natbib}
\usepackage{epsfig,amsmath,amssymb}
\usepackage{graphicx}
\usepackage{tabularx}
\usepackage{array}
\usepackage{xcolor}
\usepackage[caption=false]{subfig}
\usepackage{bm}

\def\CII{[C\,\textsc{ii}]}
\def\logNHI[#1]{$\log(N_{\rm H\scriptscriptstyle{I}}/{ \rm cm^{-2}})$ = #1}
\def\kms{km~s$^{-1}$}

\def\vsigrat{$v_{\rm rot} / \sigma_v$}

\shorttitle{The Kinematics of $z \gtrsim 6$ Quasar Host Galaxies} 
\shortauthors{Neeleman et al.}
 
\begin{document}

\title{\Large The Kinematics of $\bm{z \gtrsim 6}$ Quasar Host Galaxies}

\correspondingauthor{Marcel Neeleman}
\email{neeleman@mpia.de}

\author[0000-0002-9838-8191]{Marcel Neeleman}
\affiliation{Max-Planck-Institut f\"{u}r Astronomie, K\"{o}nigstuhl 17, D-69117, Heidelberg, Germany}
\author[0000-0001-8695-825X]{Mladen Novak}
\affiliation{Max-Planck-Institut f\"{u}r Astronomie, K\"{o}nigstuhl 17, D-69117, Heidelberg, Germany}
\author[0000-0001-9024-8322]{Bram P. Venemans}
\affiliation{Max-Planck-Institut f\"{u}r Astronomie, K\"{o}nigstuhl 17, D-69117, Heidelberg, Germany}
\author[0000-0003-4793-7880]{Fabian Walter}
\affiliation{Max-Planck-Institut f\"{u}r Astronomie, K\"{o}nigstuhl 17, D-69117, Heidelberg, Germany}
\affiliation{National Radio Astronomy Observatory, Socorro, NM 87801, USA}
\author[0000-0002-2662-8803]{Roberto Decarli}
\affiliation{INAF -- Osservatorio di Astrofisica e Scienza dello Spazio di Bologna, via Gobetti 93/3, I-40129, Bologna, Italy}
\author[0000-0002-1173-2579]{Melanie Kaasinen}
\affiliation{Max-Planck-Institut f\"{u}r Astronomie, K\"{o}nigstuhl 17, D-69117, Heidelberg, Germany}
\author[0000-0002-4544-8242]{Jan-Torge Schindler}
\affiliation{Max Planck Institut f\"ur Astronomie, K\"onigstuhl 17, D-69117, Heidelberg, Germany}
\author[0000-0002-2931-7824]{Eduardo Ba\~{n}ados}
\affiliation{Max-Planck-Institut f\"{u}r Astronomie, K\"{o}nigstuhl 17, D-69117, Heidelberg, Germany}
\author[0000-0001-6647-3861]{Chris L. Carilli}
\affiliation{National Radio Astronomy Observatory, Socorro, NM 87801, USA}
\author[0000-0002-0174-3362]{Alyssa B. Drake}
\affiliation{Max-Planck-Institut f\"{u}r Astronomie, K\"{o}nigstuhl 17, D-69117, Heidelberg, Germany}
\author[0000-0003-3310-0131]{Xiaohui Fan}
\affiliation{Steward Observatory, University of Arizona, 933 North Cherry Avenue, Tucson, AZ 85721, USA}
\author[0000-0003-4996-9069]{Hans-Walter Rix}
\affiliation{Max-Planck-Institut f\"{u}r Astronomie, K\"{o}nigstuhl 17, D-69117, Heidelberg, Germany}

\begin{abstract}
We explore the kinematics of 27 $z \gtrsim 6$ quasar host galaxies observed in \CII-158$\mu$m (\CII) emission with the Atacama Large Millimeter/sub-millimeter Array at a resolution of $\approx$\,$0\farcs25$. We find that nine of the galaxies show disturbed \CII\ emission, either due to a close companion galaxy or recent merger. Ten galaxies have smooth velocity gradients consistent with the emission arising from a gaseous disk. The remaining eight quasar host galaxies show no velocity gradient, suggesting that the gas in these systems is dispersion-dominated. All galaxies show high velocity dispersions with a mean of $129 \pm 10$ \kms. To provide an estimate of the dynamical mass within twice the half-light radius of the quasar host galaxy, we model the kinematics of the \CII\ emission line using our publicly available kinematic fitting code, \texttt{qubefit}. This results in a mean dynamical mass of $5.0\,\pm\,0.8\,(\pm\,3.5) \times 10^{10} M_\odot$. Comparison between the dynamical mass and the mass of the supermassive black hole reveals that the sample falls above the locally derived bulge mass--black hole mass relation at 2.4$\sigma$ significance. This result is robust even if we account for the large systematic uncertainties. Using several different estimators for the molecular mass, we estimate a gas mass fraction of $>$10\,\%, indicating gas makes up a large fraction of the baryonic mass of $z \gtrsim 6$ quasar host galaxies. Finally, we speculate that the large variety in \CII\ kinematics is an indication that gas accretion onto $z \gtrsim 6$ super massive black holes is not caused by a single precipitating factor.
\end{abstract}


\section{Introduction}
\label{sec:intro}

Since the first detections over half a century ago \citep{Schmidt1963, Matthews1963}, quasars have been used as beacons for detecting distant galaxies. Powered by gas accretion onto supermassive black holes (SMBHs), they are one of the most luminous non-transient objects in the Universe, and can be detected well into the epoch of reionization \citep[e.g.,][]{Mortlock2011, Banados2018}. Large optical and near-infrared sky surveys have now discovered several hundreds of quasars at $z \gtrsim 6$ \citep[e.g.,][]{Fan2000, Banados2016, Jiang2016}. 

At low redshift, studies have shown that quasars can arise in galaxies with a wide range of physical and morphological properties \citep[e.g.,][]{Bahcall1997}. However, a similar classification at high redshift is challenging, as detecting the host galaxies of $z \gtrsim 6$ quasars has proven difficult in the optical and near-infrared \citep[e.g.,][]{Mechtley2012, Decarli2012, Marshall2020a}. Instead, previous works have turned to numerical simulations, which predict that mergers are the primary cause for luminous quasars, especially at high redshifts \citep{Li2007, Hopkins2008, Capelo2015}. These simulations further predict that only in the latter stages of the merger, the conditions within the galaxy are most favorable for a SMBH to be seen as an optically luminous quasar. This quasar phase is relatively short-lived, suggesting that most high redshift quasar host galaxies should show signs of mergers, although tidal features are short-lived and could be too faint to observe \citep{Hopkins2008}.

Besides classification of the quasar host galaxies, detection of the hosts of $z \gtrsim 6$ quasars would also enable a study of the correlations between the properties of the host galaxy and the properties of the SMBH, where the latter are obtained from spectroscopy of the quasar emission lines \citep[e.g.,][]{McLure2002}. One correlation in particular that is of interest, is the correlation between the mass of the SMBH and the mass of the host galaxies. In the local Universe, a tight correlation has been observed between these two quantities \citep{Ferrarese2000, Gebhardt2000,Haering2004}, and the existence of such a tight correlation between two properties of vastly different size scales suggests that SMBHs and galaxies coevolve. Although this coevolution is supported by both indirect \citep[e.g.,][]{Aird2010, Delvecchio2014}, as well as more direct observations \citep[e.g.,][]{Alexander2012}, it remains a topic of debate \citep{Kormendy2013, Yang2019}. \citet{Kormendy2013} argue that most SMBHs at low redshifts ($z \lesssim 2$) only show low-level AGN activity which is insufficient to affect the galaxy \citep[although see][for another explanation]{AnglesAlcazar2013}. Instead, the tight correlation could be the result of averaging of galaxy and black hole properties during the mergers that formed the galaxy \citep{Jahnke2011}. 

Unlike the optical and near-infrared, where the brightness of the quasar inhibits detection of the quasar host galaxy, the Atacama Large Millimeter/sub-millimeter Array (ALMA) capitalizes on the relative faintness of the quasar at millimeter wavelengths in order to search for the quasar host. Using the \CII-158$\mu$m (\CII) emission line, which arises from the fine-structure state of singly ionized carbon, the host galaxies of $z \gtrsim 6$ quasars are now routinely detected even in short observations \citep[e.g.,][]{Decarli2018}. Because neutral carbon has an ionizing potential slightly below the ionizing potential of hydrogen, singly ionized carbon can exist in gas with a wide range of physical properties, from molecular to ionized \citep[e.g.,][]{Pineda2013}. This makes it a good kinematic tracer of the interstellar medium (ISM) of the galaxy \citep{Deblok2016}. 

However, the initial observations did not resolve the $z \gtrsim 6$ quasar host galaxies, which prevented a classification of the galaxies, and from performing a detailed study of the kinematics of the galaxies. What these observations did show were massive companion galaxies surrounding a fraction of quasar host galaxies \citep{Decarli2017}. This is in-line with the predictions from numerical simulations that indicate mergers could precipitate the quasar phase of SMBHs. In addition, the width of the integrated \CII\ emission line from unresolved observations was used as an estimate for the dynamics of the system \citep[e.g.,][]{Carilli2013}. Together with the extent of the \CII\ emission, this provided a rough estimate for the dynamical mass of the host galaxies of $z \gtrsim 6$ quasars \citep[e.g.,][]{Walter2004, Wang2013, Willott2015}. Comparison between this dynamical mass and the mass of the SMBH revealed that most $z \gtrsim 6$ quasar host galaxies are less massive compared to local galaxies with similar mass SMBHs \citep[e.g.,][]{Willott2015, Venemans2016}. It is, however, important to note that several assumptions (e.g., inclination, matter distribution, gas kinematics) were made in order to derive these dynamical masses, which resulted in substantial uncertainties.

In order to better characterize the morphology and kinematics of the gas, and provide a more reliable estimate of the dynamical mass of $z \gtrsim 6$ quasar host galaxies, higher resolution imaging is required. In this manuscript, we will explore the kinematics of a sample of 27 $z \gtrsim 6$ quasar host galaxies at $<$$0\farcs25$ resolution in the \CII\ emission line. These observations resolve the quasar host galaxies, allowing us to constrain, for the first time, the morphology and extent of the cold ISM inside a representative sample of $z \gtrsim 6$ quasar hosts. With these resolved observations we can start exploring the kinematics in a similar way as has been done at lower redshifts with gas tracers such as H$\alpha$ and [\ion{O}{3}] for non-quasar hosts \citep[e.g.,][]{Turner2017, FoersterSchreiber2018}. This allows us to determine if quasars preferentially occur in galaxies with certain kinematic properties or shows hints of recent mergers, as well as provide some of the most reliable estimates for the dynamical mass of any sample of $z \gtrsim 6$ quasar host galaxies.

This paper is one in a series of three papers discussing our sample of 27 quasars. Details of the observations and data reduction procedure for the data used in this paper are given in \citet{Venemans2020}, which also compares the \CII\ emission to the rest-frame, far-infrared continuum emission. In \citet{Novak2020} the spatial and spectral extent of the \CII\ emission is discussed for our sample. This paper is structured as followed. We describe the sample that is used in this manuscript in Section \ref{sec:sample}. In Section \ref{sec:model}, we outline the kinematic modeling of the data. In Section \ref{sec:results} we analyze the results of the kinematic modeling, which are discussed in Section \ref{sec:discussion}. Throughout this manuscript we assume a standard, flat, $\Lambda$-CDM cosmology with $\Omega_\lambda = 0.7$ and $H_0 = 70~\text{km}~\text{s}^{-1}~\text{Mpc}^{-1}$.

\section{Sample Selection}
\label{sec:sample}

Determining the morphology and kinematics as well as measuring the dynamical masses of $z \gtrsim 6$ quasar host galaxies requires a resolution such that the \CII\ emission is easily spatially resolved. Previous observations have shown that the extent of the \CII\ emission from these galaxies is $\approx$$0\farcs4$ corresponding to $\approx$2 kpc \citep[e.g.,][]{Walter2009, Wang2013, Venemans2016, Decarli2018}. We have therefore obtained $<$$0\farcs25$ (which is $<$1.2 kpc at $z \approx 6$) \CII\ imaging of $z \gtrsim 6$ quasar host galaxies using ALMA over several observing programs (in particular recent program IDs: 2017.1.01301.S and 2018.1.00908.S). We supplement this data with previous data obtained by our team, as well as archival data at the same resolution. 

This results in a sample of 27 quasars, which is discussed in detail in \citet{Venemans2020}. The observations range in resolution between $0\farcs10$ and $0\farcs25$ and have root mean square (RMS) sensitivities between 0.13 and 0.54 mJy beam$^{-1}$, with an average of 0.27 mJy beam$^{-1}$, per 30 MHz channel. The channel maps for the 27 quasars surrounding the \CII\ emission line are shown in Appendix \ref{sec:channelmaps}.

We further apportion this sample into two subsamples based on the morphology of the \CII\ emission. The first subsample consists of quasar host galaxies that have a nearby companion galaxy or show complex kinematics as the result of a recent merger. We define nearby, in this case, as close enough that it is difficult to separate the \CII\ emission from the two galaxies due to extended emission or \CII-emitting gas connecting the two galaxies \citep[see e.g.,][]{Neeleman2019}. A total of nine quasars fall within this subgroup labelled as `disturbed'. The second `undisturbed' subgroup consists of the remaining 18 galaxies whose \CII\ emission shows no signs of ongoing merger activity or close companion. We note that the \CII\ emission from four of these galaxies (see Table \ref{tab:thindiskfit}) is only barely resolved, and one quasar host galaxy (P323$+$12) is only detected at very low significance.

\section{Methods: Kinematic Modeling}
\label{sec:model}

In this section, we discuss the kinematic modeling of the \CII\ emission that was used to calculate the dynamical mass. The kinematic modeling was performed with the custom python code, \texttt{qubefit}. This code was specifically developed to analyze resolved line emission observations obtained with (sub-)millimeter interferometers, such as ALMA, taking into account the finite resolution of these observations. Currently available fitting codes are described in Section \ref{sec:othercodes}, and details of the code are discussed in Section \ref{sec:code}. The different models adopted for the \CII\ emission for these $\approx 0\farcs25$ observations of $z \gtrsim 6$ quasar host galaxies are discussed in Section \ref{sec:models}.

\subsection{Current 3D Fitting Codes}
\label{sec:othercodes}

With the advent of sensitive integral field units (IFUs) on optical telescopes and (sub-)millimeter interferometers, renewed interest has been given to modeling the emission from emission lines in the `3D' data cubes these instruments provide \citep{Jozsa2007,DiTeodoro2015,Bouche2015,Westfall2019}. These data cubes, with two spatial directions and one spectral direction, provide detailed information on the kinematic properties of the gas traced by the emission line over a 2D region of the sky. In the past, because of limited computational power, direct analysis of these data cubes was cumbersome and instead analysis was performed on 2D projections of the data cube, such as the integrated line flux density and velocity field \citep[e.g., \texttt{rotcur},][]{VanAlbada1985}. Although 2D analysis yields accurate results for observations with high spatial resolution, several effects --most notably ambiguity in determining velocity fields and beam smearing-- result in large uncertainties on the derived kinematics from low resolution, low signal-to-noise (S/N) observations \citep[see the discussion in][]{DiTeodoro2015}.

With the increased performance of current computers, it has now become feasible to fit the full 3D data cube, and several packages are currently available for this task. The fitting codes \texttt{TiRiFiC} \citep{Jozsa2007}, and \texttt{$^{\rm 3D}$Barolo} \citep{DiTeodoro2015} allow the user to fit tilted-ring models to the data cube. Whereas the former allows for a wide range of individual customizations appropriate for high angular resolution observations, the latter is optimized for low-resolution observations. The fitting code \texttt{GalPaK$^{\rm 3D}$} \citep{Bouche2015} was developed for optical IFU data, although it has also been applied to interferometric data \citep[e.g.,][]{Privon2017, CalistroRivera2018}, and fits a disk galaxy assuming a user-selected intensity and velocity profile. Finally, \texttt{KinMS} \citep{Davis2013, Davis2017} was designed for interferometric observations of local galaxies by fitting detailed parametric models to high resolution data \citep[e.g.,][]{North2019}. 

All these codes have their own strengths, and provide the user with a way to model data cubes that take into account the limited resolution of the observations. However, all codes assume that the line emission, which is being modeled, arises from a disk-like structure. In the high redshift Universe, where galaxies are often believed to be interacting, and show emission that is clumpy \citep[e.g.,][]{Cowie1995} this might be too simplistic. For instance, for the $0\farcs07$ resolution observation of the $z  = 6.61$ quasar J0305$-$3150, \citet{Venemans2019} find that a disk does not accurately reproduce the observed \CII\ emission profile.

\subsection{\texttt{qubefit}}
\label{sec:code}

In order to capture the increasing complexity of far-infrared emission line observations, and to allow for maximum flexibility in emission models, we have designed the Python-based code \texttt{qubefit}\footnote{https://github.com/mneeleman/qubefit}. Its core strength is the ability to easily create different emission models and calculate the best-fit parameters and associated uncertainties for these models using a Markov Chain Monte Carlo (MCMC) approach. The code consists of four steps.

The first step in this process is for the user to define a model of the emission. Several models are predefined in the code, including all of the models used in this paper, but the modular setup of the code allows the user to define their own model with minimal adjustment to the code. The code describing the model takes the $n$ parameters, which define the model, and generates a model cube with the same size and dimensions as the data cube. Description of the models used in this paper are given in Section \ref{sec:models}. 

In the second step, the model cube is convolved with the beam and line-spread function of the instrument. For ALMA data, the line-spread function is often negligible, because the spectral channels have been averaged enough such that consecutive channels are independent of each other. This is the case in all of our observations, where the channel spacing of 30~MHz is much larger than the line-spread function of ALMA, which is 1~MHz. We model the beam with a 2D-Gaussian with size and position angle as determined from a Gaussian fit to the beam of the data cube. We note that this is an approximation in two ways: 1) The actual beam (often referred to as the `dirty' beam) is much more complicated in interferometric data, hence the need for removing the beam before analyzing the data in a process known as \textit{cleaning}. However, running the fitting code on the full data cube is too computationally expensive. We therefore run the fitting code on a small region containing the emission in the \textit{cleaned} data cube for which the beam can be approximated by a 2D Gaussian. 2) The beam varies as a function of frequency. These variations are, however, small (less than 0.1\,\%) over the velocity range considered. We do not expect that either of these approximations will significantly affect the results presented in this paper.

In the third step, the convolved model cube ($\mathcal{M}$) is compared with the data cube ($\mathcal{D}$) using a user-defined likelihood function ($\mathcal{L}$). For this paper we use a modified $\chi^2$ likelihood function of the form:
\begin{equation}
\ln \mathcal{L} = -0.5 f_{\rm beam}  \sum_i^{i=n} \frac{(\mathcal{M}_i - \mathcal{D}_i)^2}{\mathcal{R}_i^2}
\end{equation}
Here, $\mathcal{R}$ is the RMS sensitivity of the data cube, and the sum is taken over all $n$ pixels within the user-defined mask. The mask used for the analysis in this paper is generated from all pixels with at least a 3$\sigma$-significant signal, which is then expanded to contain a buffer of at least one independent measurement around the 3$\sigma$ features. This guarantees that the majority of the emission is contained within the mask, but the mask contains few excess pixels that do not constrain the model. The last factor $f_{\rm beam}$ is an adjustment factor to account for the correlated nature of interferometric data pixels. When generating the image during the \textit{cleaning} process, the full width at half maximum (FWHM) of the beam is often over-sampled at 5 or more pixels. Under the assumption that the beam can be Nyquist-sampled at approximately 2 samples per FWHM of the beam, the adjustment factor is $f_{\rm beam} =  \pi / (A_{\rm beam}\ln2)$, where $A_{\rm beam}$ is the area of the synthesized beam in pixels. We note that this is an approximation to the true correlated nature of the observations, which requires accounting for the covariance matrix describing the pixel-to-pixel correlations as well as the uncertainty in the $\chi^2$ statistic \citep{Davis2017, Smith2019}. However, for our observations this simple approach gives similar adjustment factors as the more computationally intensive approach.

In the final step, we maximize the likelihood function in order to find the best-fit parameters, as well as estimate the uncertainties on the parameters for the model. To accomplish this, we use the affine-invariant MCMC ensemble sampler, \texttt{emcee}, to sample the multidimensional parameter space \citep{ForemanMackey2013}. We assume flat, uninformed priors on all of the parameters, and run two independent ensembles using different initial values. To check for the convergence of the ensemble, we compare the ensemble both with itself and with the other ensemble. We further remove the first 30\,\% of the ensemble as a `burn-in' period, where the chain has not converged on its steady-state solution. The two chains are combined, and the results represent the full probability distribution function (PDF) of each parameter of the model. 
For those parameters that are constrained by the model, we report median values with 16$^{\rm th}$ to 84$^{\rm th}$ percentile ranges of the PDFs. Unconstrained parameters are reported with `3$\sigma$' upper and lower limits, defined such that less than 0.14 \% of the PDF is above or below this limit.

\subsection{Models}
\label{sec:models}

As shown in previous works, the \CII\ emission from $z \gtrsim 6$ quasar host galaxies is often complex, either due to mergers or possible quasar feedback \citep{Banados2019, Decarli2019, Venemans2019}. Accurately modeling this emission is beyond the scope of this work. In this paper, we wish to explore how well we can determine the overall kinematics, under the assumption that the gas is gravitationally supported. This assumption clearly does not need to hold when a galaxy is undergoing a merger. However, for consistency we will apply the assumption to all galaxies, regardless of morphology, and then explore the accuracy of this assumption based on the morphological classification of the galaxy.

Under the assumption that the \CII-emitting gas is gravitationally supported, there are two scenarios that bracket the possible dynamics of the system. Either the gas is contained to a thin rotating disk, or the gas is distributed in a halo, in which the gas moves in random directions. The first scenario is predicted by simulations \citep[e.g.,][]{Lupi2019}, as the gas quickly interacts with itself and the angular momentum of the gas forces the system to flatten into a thin rotating disk, whereas the second scenario is consistent with the stellar motions in classical bulges and most elliptical galaxies \citep[e.g.,][]{Kormendy2013}, although this scenario requires some form of energy injection to prevent the gas from settling into a disk. The simplified models for both scenarios are described here.

\subsubsection{Thin Disk Model}
\label{sec:thindiskmodel}

The thin disk model is described by nine parameters and has been previously summarized in Appendix C of \citet{Neeleman2019}. Five parameters describe the position of the disk. The right ascension ($x_{\rm c}$), declination ($y_{\rm c}$) and redshift ($z_{\rm kin}$) of the rotational center, and the position angle of the major axis ($\alpha$) and the inclination ($i$) of the disk. Two parameters describe the emission profile of the disk. The central emission, $I_0$, and the scale length of the emission, $R_{\rm D}$. Finally, the kinematic properties of the disk are described by the rotational velocity $(v_{\rm rot}$) and the velocity dispersion ($\sigma_v$). 

The emission of the disk is assumed to be axisymmetric and exponentially decreasing with distance from the center. Although in \citet{Novak2020} we find that the \CII\ emission is best described by a compact and extended component, introducing a more complicated emission profile does not improve the fit. This is because the extended component seen in \citet{Novak2020} is faint and is only clearly visible in the stacked observations. For a cylindrical coordinate system ($r$, $\phi$, $z$) centered on the disk ($x_{\rm c}$, $y_{\rm c}$) such that the axis of the rotation points along the $z$-axis, this can be written as:
\begin{equation}
I(r) = I_0 e^{-r/R_{\rm D}}.
\end{equation}
The galactocentric radius, $r$, can be calculated for an inclined disk from the plane-of-the-sky projection of the disk as \citep{Chen2005, Neeleman2016}:
\begin{equation}
r = r' \times \sqrt{1 + \sin^2(\phi' - \alpha) \tan^2(i)}.
\end{equation}
Here $r'$ and $\phi'$ are the projected coordinates for a cylindrical coordinate system ($r'$, $\phi'$, $z'$) centered on the kinematic center of the disk such that $r'$ and $\phi'$ are in the plane of the sky.

Assuming that all of the velocities are due to rotational motion within this plane, i.e., we ignore any radial motion of the gas, then the observed, projected velocities along the line-of-sight are:
\begin{equation}
v_{0, z'} = \frac{\cos(\phi'-\alpha)\sin(i)}{\sqrt{1 + \sin^2(\phi' - \alpha)\tan^2(i)}}v_{\rm rot} + v_{\rm c}(z_{\rm kin}),
\label{eq:vlos}
\end{equation}
where $v_{\rm c}(z_{\rm kin})$ is the velocity offset from the fiducial redshift of the \CII\ emission as derived in \citet{Venemans2020}. Actual velocities along the line-of-sight will be dispersed due to internal dispersion within the disk. Assuming the dispersion obeys a 1D Gaussian distribution yields:
\begin{equation}
\label{eq:idisk}
I(r', \phi', v') = I(r', \phi') e^{(v' - v_{0, z'})^2 / 2\sigma_v^2}.
\end{equation}
Equation \ref{eq:idisk} yields a unique value for each point in the 3D model cube. We note that in this equation we have explicitly kept both the velocity and dispersion constant over the full disk. As the emission is not highly resolved in any of our galaxies, this simplification is sufficient (see Section \ref{sec:radialprofiles}). We have tested more complicated velocity and dispersion profiles, but these do not change our conclusions.

\subsubsection{Dispersion-Dominated Bulge Model}
\label{sec:bulgemodel}

For the dispersion-dominated bulge model, we assume that the \CII-emitting gas does not show any systemic rotation, and that the dynamics of the gas are described by randomly oriented motions. We can see from the \CII\ velocity fields that this may be a poor approximation for some systems that show a clear velocity gradient across the galaxy. However, we can use this model to place an upper limit on the dispersion. In total, the bulge model is described by six parameters. Three parameters describe the position of the center of the bulge: right ascension ($x_{\rm c}$), declination ($y_{\rm c}$) and redshift ($z_{\rm kin}$), two parameters describe the emission profile ($I_0$, and $R_{\rm D}$), and one parameter, the velocity dispersion ($\sigma_v$), is needed for the kinematic properties. As with the thin disk model, we assume the velocity dispersion profile is constant over the full extent of the emission. This simplification is sufficient at the current resolution.

For the emission profile, we assume that the integrated flux along the sightline is exponential, such that:
\begin{equation}
I(r') = I_0 e^{-r'/R_{\rm D}},
\end{equation}
where $r'$ is the radial distance from the kinematic center of the bulge ($x_c$, $y_c$). Note that we here have made the substitution that $r = r'$, which is valid for the spherically symmetric bulge model. The mean velocity across the bulge is assumed to be at a constant, systemic velocity ($v_{\rm c}$), with a constant, Gaussian velocity dispersion ($\sigma_v$) resulting in:
\begin{equation}
I(r', v') = I_0 e^{-r'/R_{\rm D}} e^{(v' - v_{\rm c})^2 / 2\sigma_v^2}.
\end{equation}
This yields a unique value for each point in the 3D model cube. These models are then used in the fitting routine to estimate the best-fit parameters for the model.

\subsection{Goodness of Fit}
\label{sec:gof}

To provide a measure of the goodness of fit, we calculate two goodness-of-fit statistics: a simple reduced-$\chi^2$ statistic and a one-sample Kolmogorov-Smirnov (KS) probability. Both statistics are calculated for the pixels contained within the union of the fitting mask as described in Section \ref{sec:code} and the 1$\sigma$ contour of the model. The latter is included to account for the possible over-subtraction of pixels outside the fitting mask. To mitigate the effects of correlated noise in the data due to oversampling of the beam, we only take one measurement per beam. This guarantees that the measurements are independent \citep{Condon1997}.

The reduced-$\chi^2$ measurement is calculated using the standard formula:
\begin{equation}
\text{red.-}\chi^2 = \frac{1}{n-m} \sum_i^{i=n}\frac{(\mathcal{M}_i - \mathcal{D}_i)^2}{\mathcal{R}_i^2},
\end{equation}
where $m$ is number of parameters of the model and the sum is over all $n$ independent measurements.  

The one-sample KS-statistic is calculated by comparing the residuals for the $n$ independent measurements to a Gaussian distribution with the observed RMS of the data cube. If the model is good, the residuals should have the same Gaussian noise distribution as the rest of the data cube. The KS-statistic is converted to a value between 0 and 1, using standard recipes within \texttt{scipy}, describing the probability that the residual sample was drawn from the Gaussian distribution. For instance, a value of 0.05 would indicate that we can reject the null hypothesis (i.e., that the sample was drawn from the Gaussian noise distribution) at the 95\,\% confidence level.

\subsection{Fitting Caveats}
\label{sec:fittingcaveats}

\begin{deluxetable*}{llllllllllll}
\tabletypesize{\scriptsize}
\tablecaption{Parameters of the kinematic modeling with a thin disk model
\label{tab:thindiskfit}}
\tablehead{
\colhead{Name} &
\colhead{R.A. ($x_c$)} &
\colhead{Decl. ($y_c$)} &
\colhead{$z_{\rm kin}$} &
\colhead{$\alpha$} &
\colhead{$i$} &
\colhead{$v_{\rm rot}$} &
\colhead{$\sigma_v$} &
\colhead{$I_0$} &
\colhead{$R_{\rm d}$} & 
\colhead{red.-$\chi^2$} & 
\colhead{KS prob.} \\
\colhead{} &
\colhead{(ICRS)} &
\colhead{(ICRS)} &
\colhead{} &
\colhead{($\degr$)} &
\colhead{($\degr$)} &
\colhead{(km~s$^{-1}$)} &
\colhead{(km~s$^{-1}$)} &
\colhead{(mJy~kpc$^{-2}$)} &
\colhead{(kpc)} &
\colhead{} &
\colhead{}
}
\startdata
P007+04 & 00:28:06.56548(23) & $+$04:57:25.415(3) & 6.00147(10) & 336$_{-13}^{+14}$ & $<$40 & $>$54 & \phm{$>$}148$_{-5}^{+6}$ & 21.6$_{-  1.5}^{+  1.7}$ & 0.412(20) & 0.981 & 0.748\\
J0100+2802 & 01:00:13.0222(5) & $+$28:02:25.832(6) & 6.32685(17) & 119$_{-5}^{+4}$ & \phm{$>$}53$_{-4}^{+3}$ & $<$59 & \phm{$>$}177$_{-11}^{+13}$ & 1.81 $\pm$   0.11 & 1.18 $\pm$   0.08 & 1.148 & 0.829\\
J0109-3047 & 01:09:53.12571(28) & $-$30:47:26.331(4) & 6.79028(11) & 52$_{-7}^{+11}$ & \phm{$>$}47$_{-9}^{+5}$ & $<$78 & \phm{$>$}119 $\pm$ 6 & 6.2$_{-  0.5}^{+  0.7}$ & 0.45$_{-  0.04}^{+  0.03}$ & 1.189 & 0.846\\
J0129-0035 & 01:29:58.51365(12) & $+$00:35:39.8260(25) & 5.778779(31) & 287.5 $\pm$   1.7 & \phm{$>$}11.6$_{-  2.9}^{+  3.9}$ & \phm{$>$}340$_{-90}^{+110}$ & \phm{$>$}60.1 $\pm$   1.5 & 17.0 $\pm$   0.6 & 0.539(11) & 1.046 & 0.511\\
J025-33 & 01:42:43.72308(18) & $-$33:27:45.5050(22) & 6.33738(4) & 81.5 $\pm$   1.7 & \phm{$>$}41.5$_{-  2.0}^{+  1.9}$ & \phm{$>$}141$_{-6}^{+7}$ & \phm{$>$}109.2 $\pm$   2.1 & 17.7$_{-  0.5}^{+  0.6}$ & 0.777(18) & 1.235 & 0.380\\
P036+03 & 02:26:01.87639(13) & $+$03:02:59.2483(19) & 6.54063(3) & 189.9$_{-  2.0}^{+  1.8}$ & \phm{$>$}21$_{-4}^{+5}$ & \phm{$>$}200$_{-30}^{+50}$ & \phm{$>$}62.3 $\pm$   1.2 & 2.69 $\pm$   0.07 & 0.715(17) & 1.891 & 0.023\\
J0842+1218$^a$ & 08:42:29.4323(5) & $+$12:18:50.497(9) & 6.07480(28) & 154 $\pm$ 10 & \phm{$>$}54$_{-15}^{+10}$ & \phm{$>$}150$_{-40}^{+50}$ & \phm{$>$}142$_{-18}^{+16}$ & 18$_{-4}^{+6}$ & 0.4 $\pm$   0.06 & 1.158 & 0.589\\
J1044-0125$^a$ & 10:44:33.0426(10) & $-$01:25:02.064(13) & 5.7845(9) & 295$_{-16}^{+19}$ & \phm{$>$}37$_{-17}^{+15}$ & \phm{$>$}320$_{-90}^{+240}$ & \phm{$>$}184$_{-21}^{+27}$ & 5.1$_{-  0.7}^{+  0.8}$ & 0.62$_{-  0.08}^{+  0.10}$ & 0.922 & 0.797\\
J1048-0109 & 10:48:19.0786(2) & $-$01:09:40.420(5) & 6.67578(12) & 21 $\pm$ 5 & \phm{$>$}26 $\pm$ 9 & \phm{$>$}240$_{-60}^{+100}$ & \phm{$>$}103$_{-6}^{+5}$ & 17.7$_{-  1.3}^{+  1.5}$ & 0.555(32) & 0.988 & 0.818\\
J1120+0641 & 11:20:01.4670(4) & $+$06:41:23.866(5) & 7.08498(21) & 250$_{-15}^{+16}$ & $<$38 & $>$35 & \phm{$>$}201$_{-11}^{+13}$ & 3.41$_{-  0.25}^{+  0.27}$ & 0.57 $\pm$   0.03 & 1.141 & 0.702\\
P183+05 & 12:12:26.9754(4) & $+$05:05:33.566(5) & 6.43855(10) & 260.2$_{-  2.6}^{+  2.7}$ & $<$22 & $>$320 & \phm{$>$}140 $\pm$ 5 & 10.5 $\pm$   0.4 & 1.15 $\pm$   0.03 & 1.175 & 0.741\\
P231-20 & 15:26:37.83731(16) & $-$20:50:00.7383(16) & 6.58676(6) & 83 $\pm$ 4 & \phm{$>$}45 $\pm$ 3 & \phm{$>$}42 $\pm$ 9 & \phm{$>$}119.4$_{-  2.8}^{+  3.0}$ & 7.4 $\pm$   0.4 & 0.356(16) & 1.186 & 0.808\\
J2054-0005 & 20:54:06.49920(13) & $+$00:05:14.4630(22) & 6.03900(4) & 127.4$_{-  2.2}^{+  2.1}$ & $<$18 & $>$250 & \phm{$>$}80.6$_{-  2.0}^{+  2.1}$ & 4.39$_{-  0.16}^{+  0.17}$ & 0.500(12) & 1.240 & 0.510\\
J2100-1715$^a$ & 21:00:54.6971(10) & $-$17:15:21.981(9) & 6.0808(4) & 268$_{-6}^{+7}$ & \phm{$>$}58$_{-8}^{+6}$ & \phm{$>$}80$_{-30}^{+40}$ & \phm{$>$}167$_{-20}^{+29}$ & 5.7$_{-  0.8}^{+  0.9}$ & 0.91$_{-  0.12}^{+  0.15}$ & 0.995 & 0.843\\
P323+12$^a$ & 21:32:33.1826(10) & $+$12:17:55.118(10) & 6.58758(22) & 238.6$_{-  2.9}^{+  3.3}$ & \phm{$>$}66.9$_{-  3.0}^{+  2.7}$ & $<$73 & \phm{$>$}126$_{-8}^{+10}$ & 0.272(26) & 1.14 $\pm$   0.09 & 0.968 & 0.795\\
J2318-3029 & 23:18:33.09827(20) & $-$30:29:33.593(3) & 6.14581(10) & 88 $\pm$ 3 & \phm{$>$}15$_{-5}^{+6}$ & \phm{$>$}410$_{-110}^{+180}$ & \phm{$>$}91$_{-5}^{+6}$ & 1.04$_{-  0.07}^{+  0.08}$ & 0.480(25) & 0.937 & 0.813\\
J2348-3054$^a$ & 23:48:33.34691(26) & $-$30:54:10.264(4) & 6.8999(5) & 258$_{-13}^{+11}$ & $<$65 & $>$150 & \phm{$>$}189$_{-19}^{+22}$ & 6.2$_{-  0.9}^{+  1.1}$ & 0.235(22) & 0.973 & 0.857\\
P359-06 & 23:56:32.4409(2) & $-$06:22:59.257(6) & 6.17189(10) & 315.7 $\pm$   2.6 & \phm{$>$}54.0$_{-  2.5}^{+  2.3}$ & \phm{$>$}126 $\pm$ 8 & \phm{$>$}112$_{-4}^{+5}$ & 11.0$_{-  0.6}^{+  0.7}$ & 0.97 $\pm$   0.04 & 1.200 & 0.835\\
\hline
P009-10 & 00:38:56.5230(9) & $-$10:25:54.013(12) & 6.00436(17) & 316.4 $\pm$   2.8 & \phm{$>$}60.7$_{-  2.8}^{+  2.5}$ & \phm{$>$}125 $\pm$ 12 & \phm{$>$}148$_{-8}^{+9}$ & 12.9 $\pm$   0.8 & 2.12$_{-  0.14}^{+  0.15}$ & 1.339 & 0.604\\
J0305-3150 & 03:05:16.9246(2) & $-$31:50:55.961(4) & 6.61403(5) & 86 $\pm$ 3 & \phm{$>$}38 $\pm$ 3 & \phm{$>$}86$_{-7}^{+8}$ & \phm{$>$}88.2$_{-  2.2}^{+  2.4}$ & 4.06 $\pm$   0.14 & 1.14 $\pm$   0.04 & 1.417 & 0.425\\
P065-26 & 04:21:38.0507(74) & $-$26:57:15.554(63) & 6.1894(38) & --- & $<$65 & \phm{$>$}--- & $>$170 & 2.19$_{-  0.25}^{+  0.30}$ & 2.0$_{-  0.3}^{+  0.5}$ & 1.345 & 0.728\\
P167-13 & 11:10:34.0255(4) & $-$13:29:46.534(3) & 6.51519(6) & 296.4 $\pm$   1.5 & \phm{$>$}64.7$_{-  1.4}^{+  1.3}$ & \phm{$>$}70 $\pm$ 6 & \phm{$>$}145 $\pm$ 3 & 11.7$_{-  0.5}^{+  0.6}$ & 1.13 $\pm$   0.04 & 2.046 & 0.018\\
J1306+0356 & 13:06:08.2608(7) & $+$03:56:26.247(7) & 6.03308(12) & 89 $\pm$ 6 & \phm{$>$}53$_{-6}^{+5}$ & \phm{$>$}58 $\pm$ 14 & \phm{$>$}112$_{-6}^{+7}$ & 10.5$_{-  1.0}^{+  1.2}$ & 0.98$_{-  0.09}^{+  0.10}$ & 0.948 & 0.911\\
J1319+0950 & 13:19:11.28639(30) & $+$09:50:51.497(8) & 6.13314(9) & 240.3 $\pm$   1.5 & \phm{$>$}39$_{-4}^{+3}$ & \phm{$>$}365$_{-21}^{+28}$ & \phm{$>$}77$_{-4}^{+5}$ & 12.1 $\pm$   0.9 & 1.03 $\pm$   0.04 & 1.172 & 0.753\\
J1342+0928 & 13:42:08.0998(7) & $+$09:28:38.474(21) & 7.54104(14) & 15 $\pm$ 4 & \phm{$>$}59 $\pm$ 3 & $<$11 & \phm{$>$}99$_{-6}^{+7}$ & 0.45$_{-  0.03}^{+  0.04}$ & 1.47 $\pm$   0.09 & 1.067 & 0.750\\
P308-21 & 20:32:09.9970(14) & $-$21:14:02.618(31) & 6.2295(8) & 10.4$_{-  1.6}^{+  1.4}$ & \phm{$>$}77.0$_{-  1.4}^{+  1.0}$ & \phm{$>$}440$_{-40}^{+30}$ & \phm{$>$}173$_{-11}^{+12}$ & 4.9$_{-  0.3}^{+  0.4}$ & 3.75$_{-  0.32}^{+  0.28}$ & 1.554 & 0.242\\
J2318-3113 & 23:18:18.3570(18) & $-$31:13:46.399(16) & 6.4442(5) & 251$_{-5}^{+6}$ & \phm{$>$}63$_{-6}^{+5}$ & \phm{$>$}110$_{-40}^{+70}$ & \phm{$>$}195$_{-29}^{+62}$ & 2.69$_{-  0.29}^{+  0.32}$ & 2.13$_{-  0.26}^{+  0.33}$ & 0.958 & 0.802\\
\enddata
\tablecomments{The table is apportioned by a horizontal line into two subgroups based on the morphology of the \CII\ emission. Those above the horizontal line show undisturbed \CII\ morphology, whereas those below the line show a distrubed \CII\ morphology from either a close companion or possible sign of merger activity. Parameters in this table are described in the text (Section \ref{sec:thindiskmodel}). $(a)$ Galaxy is only barely resolved in the observations and therefore kinematic modeling is difficult. These systems are also not included in the velocity and dispersion profile determination (Section \ref{sec:radialprofiles})}
\end{deluxetable*}
\begin{deluxetable*}{lllllllll}
\tabletypesize{\footnotesize}
\tablecaption{Parameters of the kinematic modeling with a dispersion-dominated bulge model
\label{tab:bulgefit}}
\tablehead{
\colhead{Name} &
\colhead{R.A. ($x_c$)} &
\colhead{Decl. ($y_c$)} &
\colhead{$z_{\rm kin}$} &
\colhead{$\sigma_v$} &
\colhead{$I_0$} &
\colhead{$R_{\rm d}$} & 
\colhead{red.-$\chi^2$} & 
\colhead{KS prob.} \\
\colhead{} &
\colhead{(ICRS)} &
\colhead{(ICRS)} &
\colhead{} &
\colhead{(km~s$^{-1}$)} &
\colhead{(mJy~kpc$^{-2}$)} &
\colhead{(kpc)} &
\colhead{} &
\colhead{}
}
\startdata
P007+04 & 00:28:06.56550(23) & $+$04:57:25.414(3) & 6.00150(10) & \phm{$>$}151$_{-5}^{+6}$ & 20.8$_{-  1.4}^{+  1.5}$ & 0.410(19) & 1.015 & 0.745\\
J0100+2802 & 01:00:13.0221(5) & $+$28:02:25.830(7) & 6.32681(16) & \phm{$>$}175$_{-11}^{+13}$ & 1.68$_{-  0.09}^{+  0.10}$ & 0.95 $\pm$   0.05 & 1.202 & 0.725\\
J0109-3047 & 01:09:53.12564(25) & $-$30:47:26.332(4) & 6.79031(10) & \phm{$>$}124 $\pm$ 5 & 6.8 $\pm$   0.5 & 0.379(19) & 1.081 & 0.879\\
J0129-0035 & 01:29:58.51382(17) & $+$00:35:39.8277(25) & 5.77875(3) & \phm{$>$}85.8 $\pm$   1.6 & 11.5 $\pm$   0.4 & 0.570(13) & 2.231 & 0.217\\
J025-33 & 01:42:43.72289(18) & $-$33:27:45.5056(23) & 6.33741(5) & \phm{$>$}134.9$_{-  2.2}^{+  2.3}$ & 14.3 $\pm$   0.4 & 0.687(14) & 2.123 & 0.025\\
P036+03 & 02:26:01.87623(17) & $+$03:02:59.2446(28) & 6.54066(4) & \phm{$>$}95.7$_{-  2.0}^{+  2.1}$ & 1.78 $\pm$   0.05 & 0.768(18) & 3.442 & 0.000\\
J0842+1218 & 08:42:29.4323(5) & $+$12:18:50.502(9) & 6.07484(25) & \phm{$>$}162$_{-11}^{+14}$ & 13.3$_{-  2.6}^{+  3.8}$ & 0.33 $\pm$   0.05 & 1.195 & 0.528\\
J1044-0125 & 10:44:33.0410(5) & $-$01:25:02.052(7) & 5.7845(11) & \phm{$>$}260$_{-30}^{+60}$ & 3.9$_{-  0.4}^{+  0.5}$ & 0.55 $\pm$   0.05 & 0.978 & 0.828\\
J1048-0109 & 10:48:19.0786(2) & $-$01:09:40.417(5) & 6.67576(12) & \phm{$>$}140$_{-5}^{+6}$ & 13.0$_{-  0.8}^{+  0.9}$ & 0.548(25) & 1.317 & 0.617\\
J1120+0641 & 11:20:01.4670(4) & $+$06:41:23.865(5) & 7.08495(21) & \phm{$>$}207$_{-11}^{+13}$ & 3.26$_{-  0.23}^{+  0.25}$ & 0.567(31) & 1.159 & 0.665\\
P183+05 & 12:12:26.9758(4) & $+$05:05:33.568(5) & 6.43852(11) & \phm{$>$}188 $\pm$ 7 & 8.18$_{-  0.29}^{+  0.30}$ & 1.21 $\pm$   0.04 & 1.549 & 0.344\\
P231-20 & 15:26:37.83734(14) & $-$20:50:00.7383(17) & 6.58679(6) & \phm{$>$}120.5$_{-  2.9}^{+  3.0}$ & 7.4 $\pm$   0.4 & 0.294(10) & 1.280 & 0.712\\
J2054-0005 & 20:54:06.49929(14) & $+$00:05:14.4692(23) & 6.03903(5) & \phm{$>$}105.0$_{-  2.5}^{+  2.7}$ & 3.36 $\pm$   0.11 & 0.524(13) & 1.880 & 0.058\\
J2100-1715 & 21:00:54.6972(8) & $-$17:15:21.982(10) & 6.0808(4) & \phm{$>$}186$_{-24}^{+38}$ & 5.3$_{-  0.7}^{+  0.9}$ & 0.64$_{-  0.07}^{+  0.08}$ & 1.033 & 0.835\\
P323+12 & 21:32:33.1821(11) & $+$12:17:55.121(11) & 6.58737(20) & \phm{$>$}116 $\pm$ 9 & 0.255(24) & 0.73$_{-  0.04}^{+  0.05}$ & 0.971 & 0.857\\
J2318-3029 & 23:18:33.09905(25) & $-$30:29:33.592(4) & 6.14601(13) & \phm{$>$}150$_{-10}^{+12}$ & 0.65 $\pm$   0.04 & 0.555(31) & 1.334 & 0.246\\
J2348-3054 & 23:48:33.34655(23) & $-$30:54:10.263(4) & 6.9001(5) & \phm{$>$}245$_{-23}^{+30}$ & 4.7$_{-  0.6}^{+  0.7}$ & 0.229(19) & 1.053 & 0.783\\
P359-06 & 23:56:32.4404(4) & $-$06:22:59.250(5) & 6.17213(11) & \phm{$>$}149$_{-6}^{+7}$ & 8.3 $\pm$   0.4 & 0.79 $\pm$   0.03 & 1.642 & 0.131\\
\hline
P009-10 & 00:38:56.5240(8) & $-$10:25:54.010(12) & 6.00431(21) & \phm{$>$}205$_{-15}^{+18}$ & 9.2 $\pm$   0.6 & 1.77$_{-  0.10}^{+  0.11}$ & 1.566 & 0.228\\
J0305-3150 & 03:05:16.9244(2) & $-$31:50:55.962(4) & 6.61399(5) & \phm{$>$}106.1$_{-  2.6}^{+  2.8}$ & 4.02 $\pm$   0.13 & 1.084(31) & 1.609 & 0.202\\
P065-26 & 04:21:38.0483(12) & $-$26:57:15.571(19) & 6.1882(12) & $>$170 & 2.05$_{-  0.26}^{+  0.30}$ & 1.75$_{-  0.23}^{+  0.28}$ & 1.316 & 0.839\\
P167-13 & 11:10:34.0249(2) & $-$13:29:46.526(3) & 6.51526(8) & \phm{$>$}152 $\pm$ 4 & 11.1 $\pm$   0.5 & 0.723(26) & 2.567 & 0.002\\
J1306+0356 & 13:06:08.2608(5) & $+$03:56:26.249(8) & 6.03303(12) & \phm{$>$}115$_{-6}^{+7}$ & 9.9$_{-  0.9}^{+  1.1}$ & 0.77 $\pm$   0.06 & 1.001 & 0.871\\
J1319+0950 & 13:19:11.2797(4) & $+$09:50:51.451(8) & $>$6.1 & \phm{$>$}440$_{-80}^{+70}$ & 6.8 $\pm$   0.6 & 0.83 $\pm$   0.05 & 1.813 & 0.017\\
J1342+0928 & 13:42:08.0995(8) & $+$09:28:38.503(21) & 7.54091(16) & \phm{$>$}99$_{-6}^{+7}$ & 0.387(29) & 1.15 $\pm$   0.06 & 1.079 & 0.658\\
P308-21 & 20:32:10.0005(9) & $-$21:14:02.387(17) & 6.2343(15) & $>$340 & 1.82$_{-  0.09}^{+  0.10}$ & 2.79$_{-  0.21}^{+  0.23}$ & 2.069 & 0.003\\
J2318-3113 & 23:18:18.3576(14) & $-$31:13:46.397(18) & 6.4440(8) & $>$150 & 2.04$_{-  0.21}^{+  0.24}$ & 1.77$_{-  0.19}^{+  0.23}$ & 1.017 & 0.810\\
\enddata
\tablecomments{The table is apportioned by a horizontal line into two
  subgroups based on the morphology of the \CII\ emission. Those above
  the horizontal line show undisturbed \CII\ morphology, whereas those
  below the line show a distrubed \CII\ morphology from either a close
  companion or possible sign of merger activity. Parameters in this table are described in the text (Section \ref{sec:bulgemodel})}
\end{deluxetable*}

It is important to note the limitations of the fitting routines discussed here. When uncertainties are calculated on the $m$ parameters of a model through a Bayesian approach, the uncertainties that are returned are based solely on comparison with the best-fit parameters of that model. Small uncertainties on a parameter do not necessarily imply a great fit, but simply imply a steep gradient of the likelihood function along that direction in the $m$-dimensional parameter space of that model. Systematic uncertainties from the choice of the model are not included in these uncertainty estimates. 

We stress that these models are by necessity very simplified, and the \CII\ emission resulting from $z \gtrsim 6$ quasar host galaxies could be much more complicated. Although the resultant fits have, in some cases, reasonably low ($\sim$1) reduced-$\chi^2$ values and high KS probabilities, this does not imply that the \CII\ emission is constrained to a disk or bulge. It only implies that at this resolution \emph{and} sensitivity the emission can be accurately described with the given model. Higher resolution and/or more sensitive observations could --and most likely will-- reveal \CII\ kinematics that is more complex than these simple models.

\section{Results}
\label{sec:results}

\subsection{Kinematic Modeling Results}
\label{sec:modelresults}

Details of the kinematic modeling are described in Section \ref{sec:model} and a more detailed analysis of the results from the fitting procedure are described in Appendix \ref{sec:kinanalysis}. The final constraints on the parameters of the kinematics are given in Table \ref{tab:thindiskfit} for the thin disk model, and in Table \ref{tab:bulgefit} for the dispersion-dominated bulge model. Both tables have been divided into two subsamples based on the morphology of the \CII\ emission (Sec. \ref{sec:sample}). The final two columns describe the goodness of fit measures, the reduced-$\chi^2$ statistic and the KS-probability. 

We find that for all quasar host galaxies, the kinematic center of the \CII\ emission line agrees within the uncertainties with the position of the center of the \CII\ emission as determined from a 2D-Gaussian fit to the total \CII\ emission \citep{Venemans2020}. In part, this is by design, because the kinematic code fits an exponential profile in much the same way as the fitting routines used in \citet{Venemans2020}. Similarly, the redshift determined from the kinematic modeling ($z_{\rm kin}$) agrees very well with the redshift determination of the total spectrum in \citet{Venemans2020}. This is largely due to the Gaussian shape of the emission line profile (Section \ref{sec:highdisp}). 

For the thin disk model, the low reduced-$\chi^2$ values and high KS-probabilities indicate that this model sufficiently describes the bulk of the \CII\ emission for most quasar host galaxies with undisturbed \CII\ morphologies. However, some of the quasar host galaxies are only barely resolved in these observations, and have large uncertainties on the inclination of the \CII\ emission (i.e., J0842$+$1218, J1044$-$0125, J2100$-$1715, P323$+$12 and J2348$-$3054). For these galaxies it remains difficult to determine the kinematics of the galaxy. We, however, keep these objects in our sample to avoid biasing ourselves toward systems with more extended emission. The largest reduced-$\chi^2$ value is for quasar host galaxy P036$+$03. The residual channel map for this galaxy (Appendix \ref{sec:kinanalysis}) reveals that the emission is poorly approximated by an azimuthally constant Gaussian distribution, and instead varies with azimuth. This could be due to a warp in the disk or the presence of spiral arms. Higher resolution observations are needed to determine the cause.

The results for the subsample of galaxies with disturbed \CII\ emission from a merger and/or close companion are more difficult to interpret. In some cases (i.e., J1342$+$0928, P308$-$21 and J2318$-$3113), the model tries to fit multiple distinct \CII\ components with a single smooth disk. This causes unreasonably large inclinations (the median inclination is about 20$\degr$ larger) in the merger subsample, and hence results in inaccurate kinematic parameters for these galaxies. In the remaining galaxies of the subsample, the central quasar host is fitted reasonably well, but the observed extended emission cannot be modeled with this simple model. Although this results in a systematic underestimation of the flux profile, it does not significantly affect the kinematic properties of the quasar host galaxy. Nevertheless, we take a conservative approach, and only use those galaxies that are part of the undisturbed subsample to determine average galaxy properties of the $z \gtrsim 6$ sample.

In general, the dispersion-dominated bulge model provides a poorer fit to the data than the thin disk model. For the undisturbed \CII\ emission subsample, only those galaxies that showed little rotational velocity or are very compact remain well-described by this simpler model. For the disturbed \CII\ emission subsample, the fits are generally much poorer, because the observed \CII\ profiles are far from spherically symmetric. Only for J1306$+$0356, whose \CII\ emission can be separated enough from the companion and shows little rotation, and those quasar host galaxies that have limited S/N (i.e., P065$-$26, J1342$+$0928, J2318$-$3113) are the reduced-$\chi^2$ values low. We note that for all systems where the dispersion model provides a good fit, the parameters of the dispersion model are consistent with the disk profile. For the remainder of the paper, we will therefore use the results from the thin disk model.

We can compare the kinematic modeling results for those quasars that have previous kinematic results published in the literature. Here we only focus on those studies for which the resolution of the observations is comparable to ours, and the kinematic parameters were estimated from resolved observations. This yields two quasars as part of our undisturbed subsample: J0129$-$0035 and P183$+$05 \citep{Wang2019, Pensabene2020}, as well as one quasar in our disturbed subsample: J1319$+$0950 \citep{Jones2017, Shao2017, Pensabene2020}\footnote{\citet{Pensabene2020} discusses two additional quasars at $<$$0\farcs25$ but they report no kinematic parameters for these quasars because of the disturbed morphology of the \CII\ emission for these sources.}. We find that our analysis yields kinematic parameters consistent with the published results for both J0129$-$0035 and P183$+$05. For J1319$+$0950 we get kinematic properties that are consistent with \citet{Jones2017} and \citet{Shao2017}, but inconsistent with \citet{Pensabene2020}. This inconsistency is probably caused by the weak companion galaxy detected in our observations, which skews the inclination estimate if not properly taken into account.

\subsection{Mean Velocity and Velocity Dispersion Fields}
\label{sec:veldispfield}

\begin{figure*}[!b]
\includegraphics[width=\textwidth]{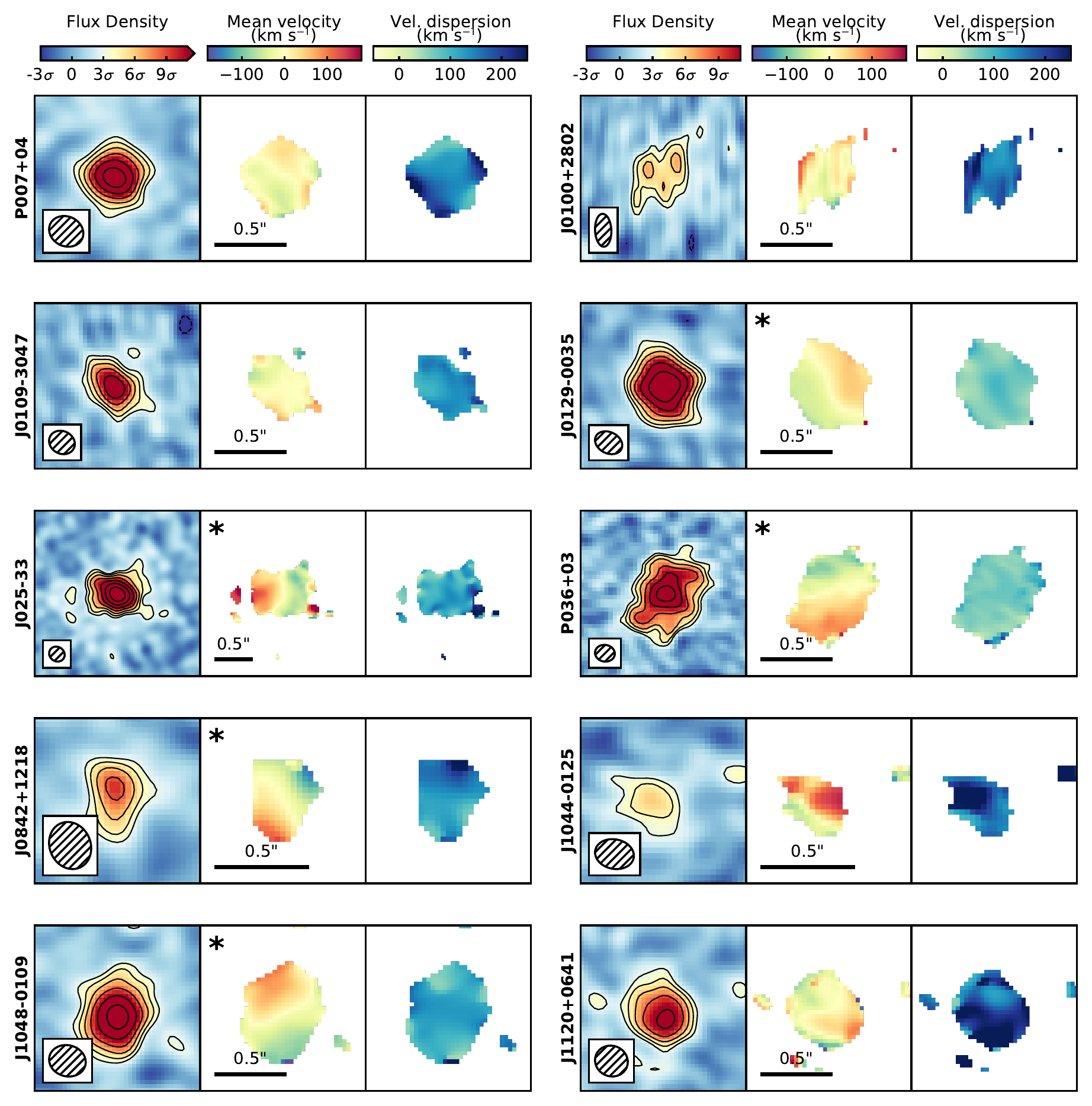}
\caption{Flux density, mean velocity and velocity dispersion maps of the 18 quasar host galaxies belonging to the `undisturbed' quasar host sample. The left panel shows the flux density of the \CII\ emission line for a channel centered on the \CII\ emission with a channel width of $1.2\times$ the FWHM \citep[see][]{Venemans2020}. The ALMA synthesized beam is shown in the bottom left inset and contours are drawn at 3$\sigma$ and increase in powers of $\sqrt{2}$ with negative contours dashed. The middle panel shows the mean velocity field, where we define the zero velocity to be at the systemic redshift of the \CII\ emission, which was determined from the integrated \CII\ spectrum \citep{Venemans2020}. The black scale bar shows an angle of $0\farcs5$, which corresponds to a physical distance of about 2.7 kpc at the redshift of the quasar. Quasar host galaxies that show a smooth  velocity gradient are marked with an asterisk. The right panel displays the velocity dispersion of \CII\ emission. For the latter two columns we only show pixels that have been detected at 3$\sigma$ in the \CII\ flux density map. In all panels north is up and east is to the left.
\label{fig:VFieldSingle}}
\end{figure*}

\begin{figure*}[!ht]\ContinuedFloat
\includegraphics[width=\textwidth]{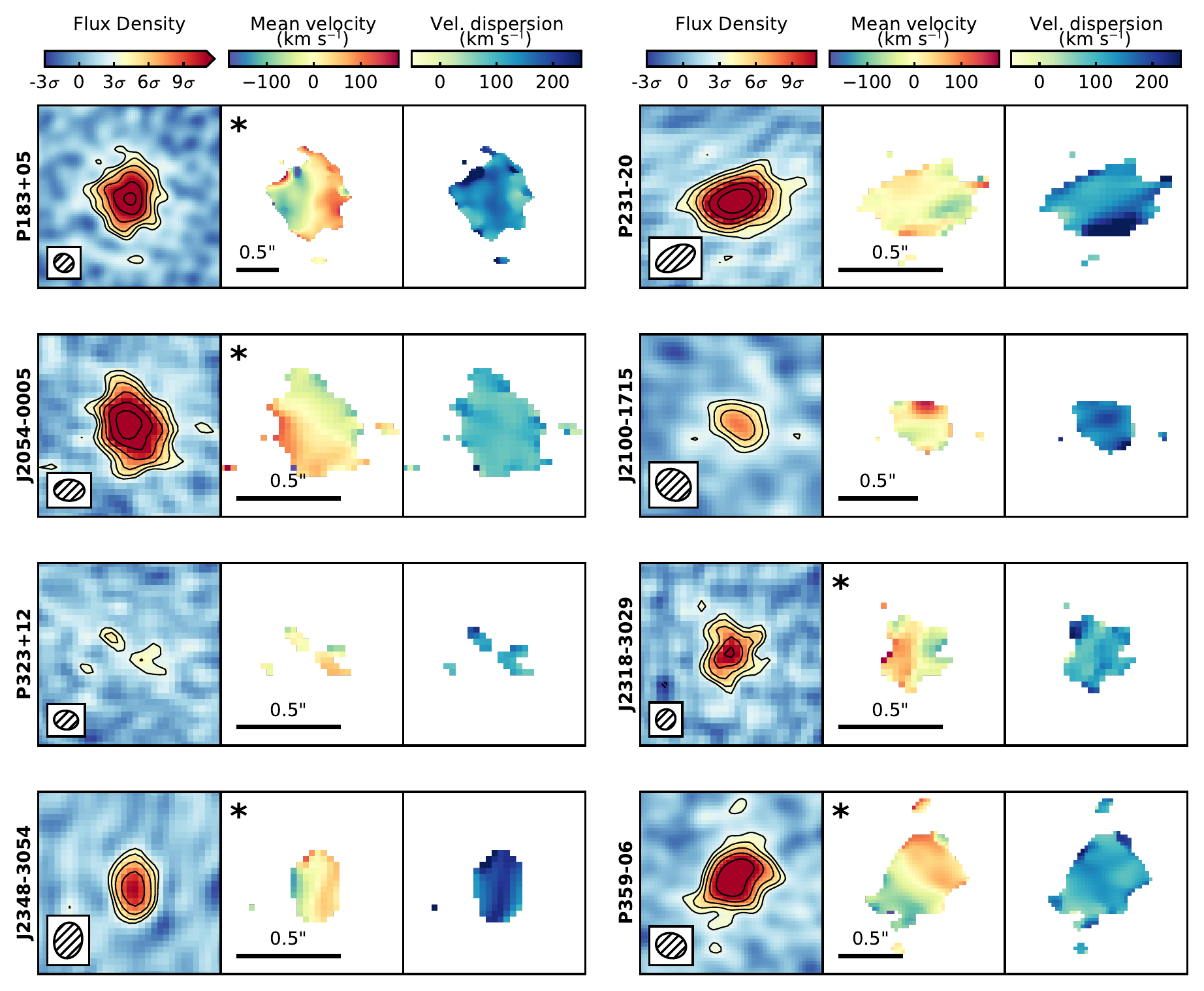}
\caption{Continued.}
\end{figure*}

\begin{figure*}
\includegraphics[width=\textwidth]{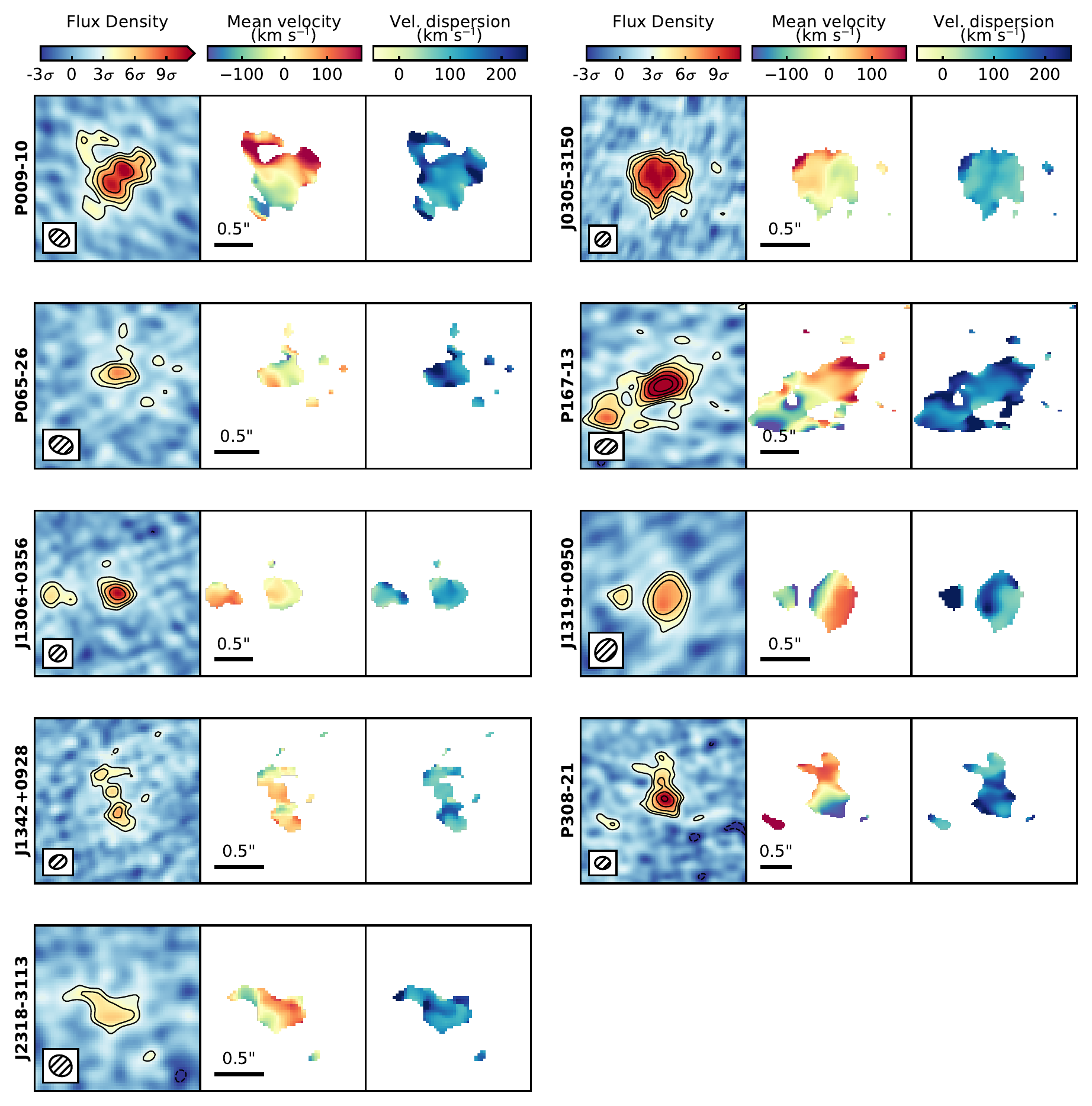}
\caption{Same as Fig. \ref{fig:VFieldSingle}, but for those QSOs that show complex \CII\ emission structures indicative of a recent merger and/or close companion galaxy.
\label{fig:VFieldComplex}}
\end{figure*}

A common way to visualize the gas kinematics from a 3D data cube is to create 2D maps of the systemic velocity and velocity dispersion of the gas at each spatial position. As mentioned in Section \ref{sec:othercodes}, there exist several methods to create these velocity fields from the data \citep[see e.g.,][]{Deblok2008}. The best method to apply depends on the resolution and S/N of the observations, as well as the intrinsic shape of the line profile. In Appendix \ref{sec:velfieldmethods}, we compare the two most common methods for deriving velocity fields; calculating the first and second moments of the data cube and Gaussian fitting of the spectra at individual spatial positions. We show that the latter method is more robust for the moderate resolution and S/N observations discussed here.

In Figure \ref{fig:VFieldSingle} and Figure \ref{fig:VFieldComplex}, we show the set of velocity fields for all 27 quasar host galaxies divided by \CII\ morphology. For the 18 quasar host galaxies in the undisturbed morphology sample, ten show a smooth velocity gradient across the \CII\ emission (Fig.~\ref{fig:VFieldSingle}). In addition, for eight of these galaxies the emission is resolved over more than two synthesized beams without showing any evidence for multiple components. The \CII\ emission is centered on the SMBH position \citep[see][]{Venemans2020}, suggesting that for these sources the \CII\ emission arises from rotating gas with the SMBH at the center of rotation.

For the remaining eight sources ---five of which are resolved over more than two synthesized beams--- the \CII\ emission is also centered around the SMBH position. However, little ordered motion is observed within these systems. One possibility is that we are observing these galaxies nearly face-on, thereby largely removing the line-of-sight velocity gradient. However, measurements of the inclination angle are consistent between these galaxies and those that show a velocity gradient (Table \ref{tab:thindiskfit}). In addition, these systems show on-average a larger velocity dispersion than galaxies that show a velocity gradient, $153 \pm 32$ compared to $95 \pm 25$~\kms. This suggests that the difference is not simply due to the viewing angle, but that the galaxy kinematics are inherently different between these two quasar host galaxy populations (Section \ref{sec:highdisp}).

The nine quasar host galaxies that show disturbed \CII\ kinematics are shown in Figure \ref{fig:VFieldComplex}. Most of the velocity fields for these quasar host galaxies have been discussed elsewhere \citep{Banados2019, Neeleman2019, Decarli2019, Venemans2019}. Some of these quasar host galaxies are marked by complex kinematics (e.g., P009$-$10, P308$-$21), whereas in others the quasar host galaxy remains relatively unaffected by the companion, because either the companion is markedly fainter (e.g., J1319$+$0950) or sufficiently far away (e.g., J1306$+$0356). For quasar host galaxies falling in the latter category, the \CII\ kinematics show both sources with strong velocity gradients and without velocity gradients similar to the galaxies in the non-disturbed subsample.

\subsection{Radial Profiles of the Mean Velocity and Velocity Dispersion}
\label{sec:radialprofiles}

\begin{figure*}
\includegraphics[width=\textwidth]{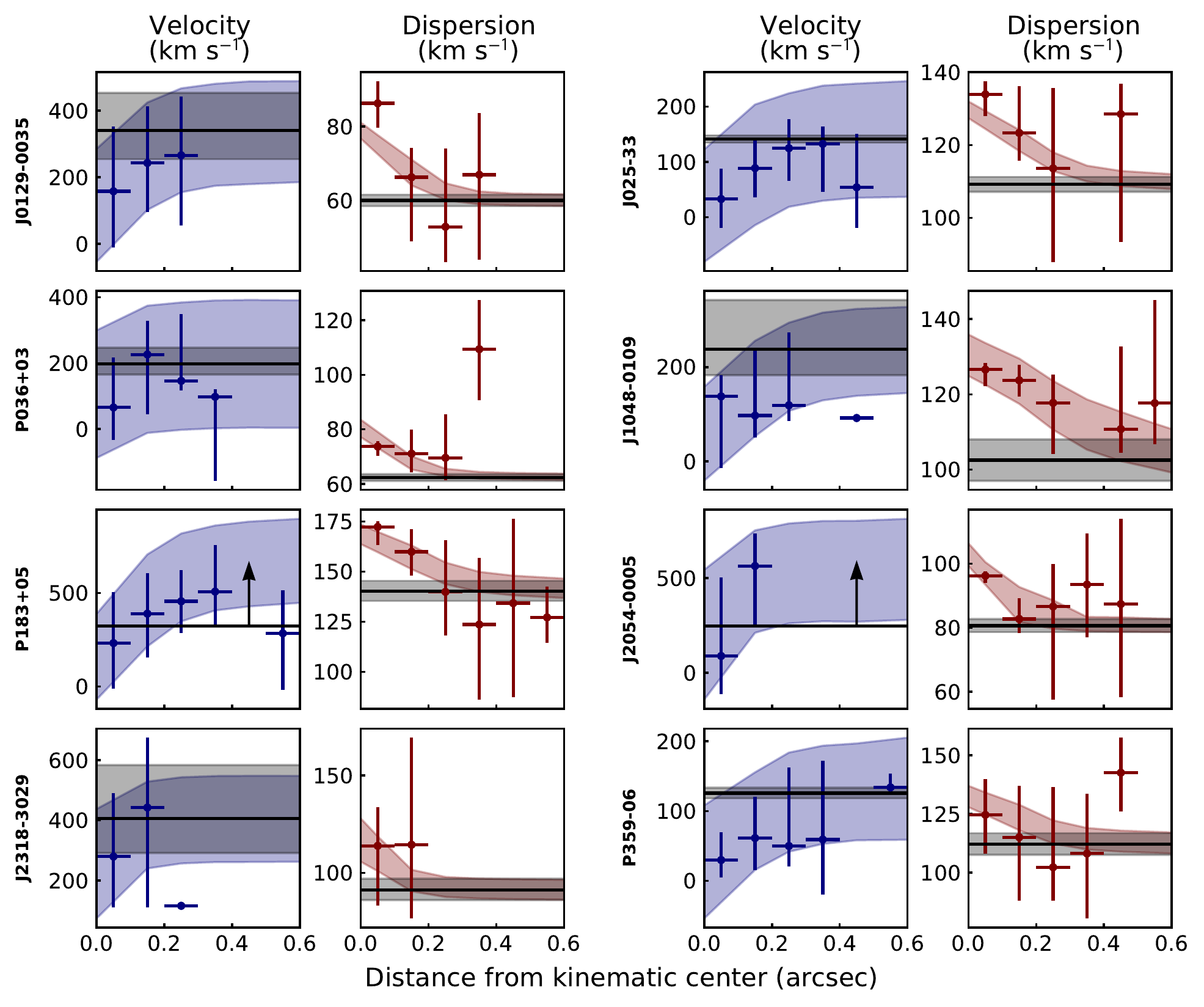}
\caption{Velocity and dispersion profiles of the eight galaxies that have emission extended over greater than two beams along the major axis, show a clear velocity gradient, and show no evidence of merger activity or nearby companion. Left panel displays the velocity profile for the data with 1$\sigma$ uncertainties. The blue shaded region marks the 16$^{\rm th}$ to 84$^{\rm th}$ percentile range of the rotational velocity profile as determined from the constant velocity model convolved with the beam.  The black (and gray shading) is the constant rotational velocity (and 1$\sigma$ uncertainties) as determined from the kinematic modeling. Right panel is similar as the left except for the velocity dispersion measurements. In all galaxies the agreement between the model and data for both the rotational velocity and velocity dispersion suggests that the assumed constant rotational velocity and velocity dispersion is sufficient at this resolution.
\label{fig:VDprofileRot}} 
\end{figure*}

\begin{figure*}
\includegraphics[width=\textwidth]{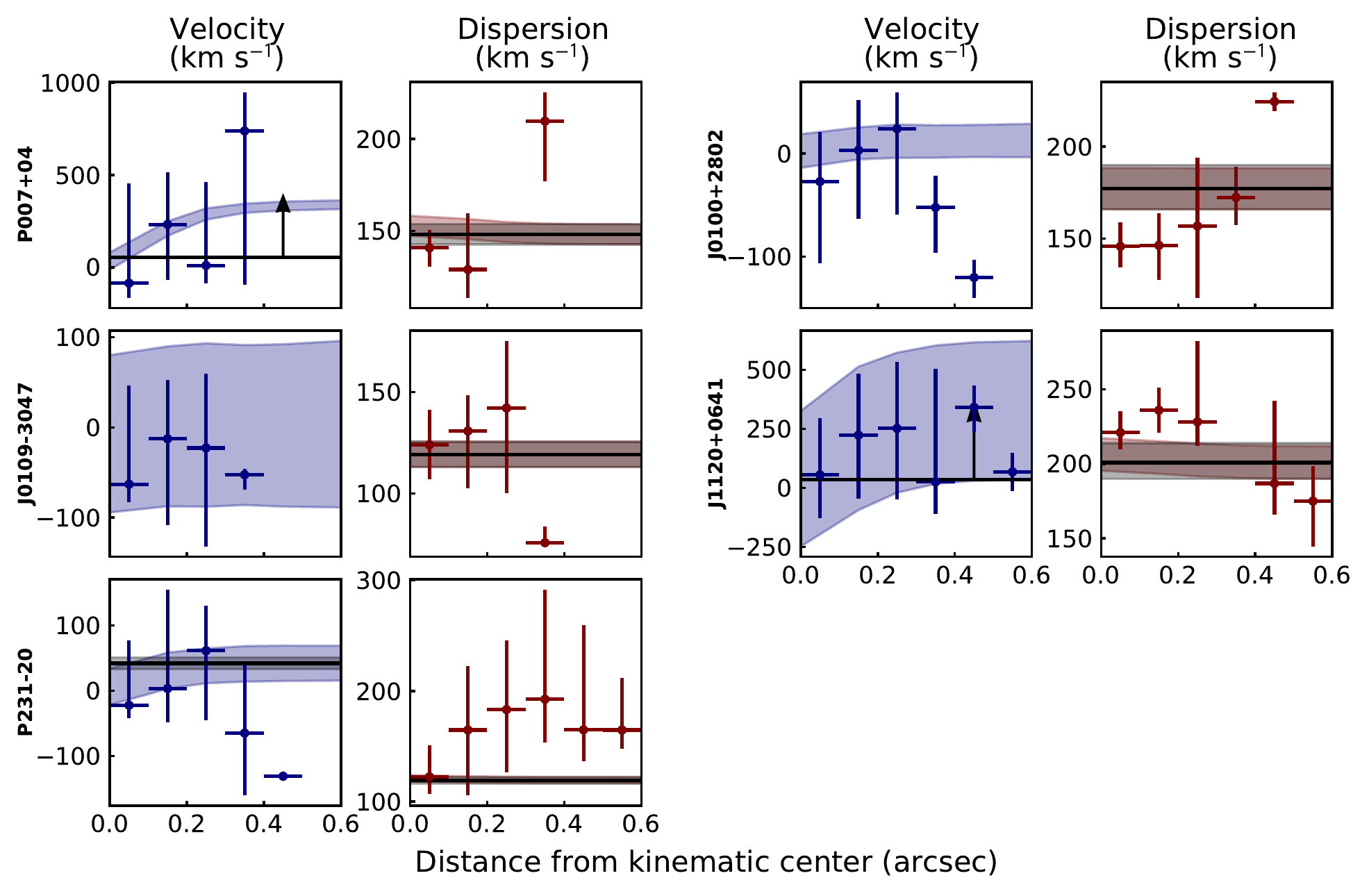}
\caption{Same as Fig. \ref{fig:VDprofileRot} for the five galaxies that have emission extended over greater than two beams, but show no signs of coherent rotation. Here the lack of a rotational gradient implies that the velocity dispersion does not increase in the center due to the effects of beam smearing. This is observed both in the data and the model, which again indicates that at this resolution the data is well-modeled by a velocity dispersion that is constant throughout the \CII\ emitting region.
\label{fig:VDprofileNR}}
\end{figure*}

With the approximately $0\farcs25$ resolution \CII\ observations, we can start exploring the radial profiles of the kinematics of the gas probed by \CII. Here we will only consider the 13 galaxies that show no signs of mergers, and are resolved over more than two beams (see Table \ref{tab:thindiskfit}). We have created the mean velocity radial profiles (also known as rotation curves) by taking the mean velocity field shown in Figure \ref{fig:VFieldSingle} and assuming that this velocity is due only to the line-of-sight velocity from motion within the plane of the galaxy disk. This allows us to compute the rotational velocity, $v_{\rm rot}$, at each pixel in the mean velocity field by inverting Equation~\ref{eq:vlos}:
\begin{equation}
v_{\rm rot} = \frac{\sqrt{1 + \sin^2(\phi' - \alpha)\tan^2(i)}}{\cos(\phi'-\alpha)\sin(i)}(v_{0, z'} - v_{\rm c}(z_{\rm kin})).
\end{equation}
As in Equation \ref{eq:vlos}, $\phi'$ is the position angle of the spatial position with respect to north and $\alpha$ and $i$ are the position angle and inclination, respectively. We also correct the line-of-sight velocity $v_{0, z'}$ to the systemic velocity of the kinematic center, $v_{\rm c}$ ($z_{\rm kin}$). We note that this formula is just a generalization of the more common approach of taking a slice of the data cube (known as a position-velocity cut) along the major axis of the galaxy, and then correcting the velocities for the inclination. Indeed the above formula reduces to the typical $1/\sin(i)$ correction for positions along the major axis of the galaxy. The individual rotational measurements are then grouped into bins based on the de-projected radius of the individual measurements. We remove measurements within 30$\degr$ of the minor axis, because uncertainties increase for these measurements due to larger velocity corrections and larger beam-smearing effects. 

The results are plotted in Figure \ref{fig:VDprofileRot} and \ref{fig:VDprofileNR}. In several cases, most notably J025$-$33 and P183$+$05, we see a general trend of increasing velocity with distance from the kinematic center of the observations. Although tempting to attribute this increase to an increasing rotation curve, it could also be the result of increased beam smearing in the central regions. To explore how much beam smearing can account for the observed rise in rotational velocity, we apply the same method to the model cube with constant velocity which has been convolved with the beam. As the figure shows, this constant velocity model shows a similar rise in the rotational velocity with radius, indicating that the increase in rotational velocity for all galaxies is largely driven by the effects of beam smearing. The resolution of these observations is therefore insufficient to measure the shape of the rotation curve, which requires at least four independent measurements along the major axis. Thus, modeling the rotational velocity with a constant function is appropriate at this resolution.

We apply the same method as described above to the velocity dispersion maps\footnote{We do not correct the 1D line-of-sight velocity dispersion measurement, to facilitate comparison with previous studies \citep[e.g.,][]{Turner2017, Hung2019, Lupi2019}. Under the assumption that the velocity dispersion is isotropic, the 3D velocity dispersion would be larger by a factor of $\sqrt{3}$.}. The results are plotted in Figure \ref{fig:VDprofileRot} and \ref{fig:VDprofileNR} for both the data and the model. Most quasar host galaxies with a strong velocity gradient have higher velocity dispersion at the center. However, this increase is also seen in their beam-convolved model data cubes, again indicating that beam-smearing is the dominant cause for this increase. This is further corroborated by the sample of quasar host galaxies without a strong velocity gradient, as these systems lack a definitive increase in their velocity dispersion in the center. At the current resolution, the assumption of a constant velocity dispersion therefore is appropriate.

We note in passing here that the current resolution remains too coarse to resolve the SMBH's sphere of influence. Therefore the constant velocity and velocity dispersion near the center of the galaxy are consistent with expectations.

\subsection{Dynamical Masses}
\label{sec:dynmass}

Having an estimation of the typical kinematics and extent of a galaxy allows us to provide a constraint on the dynamical mass of the galaxy, $M_{\rm dyn}$. Under the assumption that the mass distribution is spherically symmetric, the dynamical mass enclosed within a radius, $R$, is:
\begin{equation}
M_{\rm dyn}(R) = \frac{v_{\rm circ}^2}{G} R = 2.33 \times 10^5 v_{\rm circ}^2 R,
\label{eq:mdyn}
\end{equation}
where the latter equality holds if the circular velocity, $v_{\rm circ}$, is given in \kms, $R$ in kpc, and the dynamical mass is returned in $M_\odot$. This formula has been applied previously for $z \gtrsim 6$ quasar host galaxies \citep[e.g.,][]{Walter2004, Wang2013, Decarli2018}, but with a range of different methods for estimating the circular velocities and for different assumptions on the extent of the emission. In Section \ref{sec:rotvel} we compare the different methods for estimating the circular velocity, and in Section \ref{sec:extent} we explore how the dynamical mass estimates vary assuming different definitions of the extent of the galaxy. However, we first discuss the implicit assumption of a spherically symmetric mass distribution in using Equation \ref{eq:mdyn}.

For $z \gtrsim 6$ quasar host galaxies, we assume four main mass components that influence the kinematics within the inner few kpc of the galaxy: the SMBH, the dark matter, the stars, and the gas. For the first three mass components we assume a spherically symmetric potential. The SMBH can be considered a point source at all radii of interest, and its potential is therefore spherically symmetric by definition. Similarly, the dark matter distribution is often assumed to be spherically symmetric \citep{Navarro1997}, although it contributes little to the total mass within the inner few kpc of the galaxy \citep{Genzel2017, Price2020}. The mass distribution of stars is more difficult to estimate, as the host galaxies of $z \gtrsim 6$ quasars remain challenging to observe in the optical and near-infrared. However, simulations predict that most of the stars are within a bulge-like component that is roughly spherical \citep{Marshall2020b}, so the assumption of a spherically symmetric potential for the stellar mass distribution is probably sufficient in most galaxies.

The mass distribution of the gas in the ISM of the galaxy is probably not spherically symmetric. Simulations suggest that the cold ISM quickly settles into a cold, thin disk \citep[e.g.,][]{Lupi2019}. Although the exact mass contribution of this component to the total mass content of $z \sim 6$ quasar host galaxies is unknown, the strength of the far-infrared lines emitted from the cold ISM indicate that it could be significant (see also Section \ref{sec:MgasMdyn}). To account for the potentially non-spherically symmetric mass distribution, previous studies have calculated the velocity correction factor for a disk potential \citep{Binney2008, Walter1997}. These studies show that for all radii of interest, the dynamical mass estimate of Equation \ref{eq:mdyn} will overestimate the dynamical mass up to 30 \%, if all of the mass was constrained within a thin disk. In practice, the effective total mass distribution falls somewhere in between a thin disk and a sphere, we therefore conservatively increase the dynamical mass uncertainty by 20 \% toward lower masses to account for this systematic uncertainty.

\subsubsection{Methods for Estimating Circular Velocities}
\label{sec:rotvel}

\begin{deluxetable*}{lllllllll}
\tabletypesize{\footnotesize}
\colnumbers
\tablecaption{Circular velocity estimates\label{tab:vcirc}}
\tablehead{
\colhead{Name} &
\colhead{FWHM} &
\colhead{$i$} &
\colhead{$v_{\rm rot}$} &
\colhead{$\sigma_v$} &
\colhead{$R_{\rm e}$} &
\multicolumn{3}{c}{$v_{\rm circ}$} \\
\cline{7-9}
\colhead{} &
\colhead{} &
\colhead{} &
\colhead{} &
\colhead{} &
\colhead{} &
\colhead{0.52\,FWHM} &
\colhead{0.75\,FWHM/sin\,i} &
\colhead{$\eta=3.4$} \\
\colhead{} &
\colhead{(\kms)} &
\colhead{($\degr$)} &
\colhead{(\kms)} &
\colhead{(\kms)} &
\colhead{(kpc)} &
\colhead{(\kms)} &
\colhead{(\kms)} &
\colhead{(\kms)} 
}
\startdata
P007+04$^a$ & 370 $\pm$ 22 & $<$40 & $>$54 & \phm{$>$}148$_{-5}^{+6}$ &0.69 $\pm$ 0.03 & 192 $\pm$ 11 & $>$430 & $>$250\\
J0100+2802$^a$ & 405 $\pm$ 20 & \phm{$>$}53$_{-4}^{+3}$ & $<$59 & \phm{$>$}177$_{-11}^{+13}$ &1.98$_{-0.13}^{+0.14}$ & 211 $\pm$ 10 & \phm{$>$}382$_{-26}^{+27}$ & \phm{$>$}332$_{-21}^{+24}$\\
J0109-3047$^a$ & 354 $\pm$ 34 & \phm{$>$}47$_{-9}^{+5}$ & $<$78 & \phm{$>$}119 $\pm$ 6 &0.75$_{-0.08}^{+0.06}$ & 184 $\pm$ 18 & \phm{$>$}360$_{-50}^{+60}$ & \phm{$>$}233$_{-12}^{+11}$\\
J0129-0035 & 206 $\pm$  9 & \phm{$>$}11.6$_{-2.9}^{+3.9}$ & \phm{$>$}340$_{-90}^{+110}$ & \phm{$>$}60.1 $\pm$ 1.5 &0.904(18) & 107 $\pm$ 5 & \phm{$>$}770$_{-260}^{+200}$ & \phm{$>$}360$_{-80}^{+110}$\\
J025-33 & 370 $\pm$ 16 & \phm{$>$}41.5$_{-2.0}^{+1.9}$ & \phm{$>$}141$_{-6}^{+7}$ & \phm{$>$}109.2 $\pm$ 2.1 &1.3 $\pm$ 0.03 & 192 $\pm$ 8 & \phm{$>$}419 $\pm$ 24 & \phm{$>$}246 $\pm$ 5\\
P036+03 & 237 $\pm$  7 & \phm{$>$}21$_{-4}^{+5}$ & \phm{$>$}200$_{-30}^{+50}$ & \phm{$>$}62.3 $\pm$ 1.2 &1.199(29) & 123 $\pm$ 4 & \phm{$>$}490 $\pm$ 100 & \phm{$>$}229$_{-29}^{+42}$\\
J0842+1218 & 378 $\pm$ 52 & \phm{$>$}54$_{-15}^{+10}$ & \phm{$>$}150$_{-40}^{+50}$ & \phm{$>$}142$_{-18}^{+16}$ &0.67$_{-0.10}^{+0.11}$ & 197 $\pm$ 27 & \phm{$>$}350$_{-70}^{+80}$ & \phm{$>$}300$_{-30}^{+40}$\\
J1044-0125$^a$ & 454 $\pm$ 60 & \phm{$>$}37$_{-17}^{+15}$ & \phm{$>$}320$_{-90}^{+240}$ & \phm{$>$}184$_{-21}^{+27}$ &1.03$_{-0.13}^{+0.17}$ & 240 $\pm$ 30 & \phm{$>$}560$_{-210}^{+230}$ & \phm{$>$}460$_{-70}^{+170}$\\
J1048-0109 & 299 $\pm$ 24 & \phm{$>$}26 $\pm$ 9 & \phm{$>$}240$_{-60}^{+100}$ & \phm{$>$}103$_{-6}^{+5}$ &0.93 $\pm$ 0.05 & 155 $\pm$ 12 & \phm{$>$}500 $\pm$ 160 & \phm{$>$}300$_{-40}^{+80}$\\
J1120+0641$^a$ & 416 $\pm$ 39 & $<$38 & $>$35 & \phm{$>$}201$_{-11}^{+13}$ &0.95 $\pm$ 0.05 & 216 $\pm$ 20 & $>$510 & $>$320\\
P183+05 & 397 $\pm$ 19 & $<$22 & $>$320 & \phm{$>$}140 $\pm$ 5 &1.93$_{-0.05}^{+0.06}$ & 206 $\pm$ 10 & $>$810 & $>$400\\
P231-20$^a$ & 393 $\pm$ 35 & \phm{$>$}45 $\pm$ 3 & \phm{$>$}42 $\pm$ 9 & \phm{$>$}119.4$_{-2.8}^{+3.0}$ &0.598(26) & 204 $\pm$ 18 & \phm{$>$}410 $\pm$ 40 & \phm{$>$}224$_{-5}^{+6}$\\
J2054-0005 & 236 $\pm$ 12 & $<$18 & $>$250 & \phm{$>$}80.6$_{-2.0}^{+2.1}$ &0.839(21) & 123 $\pm$ 6 & $>$580 & $>$280\\
J2100-1715$^a$ & 361 $\pm$ 41 & \phm{$>$}58$_{-8}^{+6}$ & \phm{$>$}80$_{-30}^{+40}$ & \phm{$>$}167$_{-20}^{+29}$ &1.53$_{-0.21}^{+0.25}$ & 188 $\pm$ 21 & \phm{$>$}320 $\pm$ 40 & \phm{$>$}320$_{-40}^{+50}$\\
P323+12$^a$ & 271 $\pm$ 38 & \phm{$>$}66.9$_{-3.0}^{+2.7}$ & $<$73 & \phm{$>$}126$_{-8}^{+10}$ &1.92$_{-0.15}^{+0.16}$ & 141 $\pm$ 20 & \phm{$>$}220 $\pm$ 30 & \phm{$>$}243$_{-16}^{+18}$\\
J2318-3029 & 293 $\pm$ 17 & \phm{$>$}15$_{-5}^{+6}$ & \phm{$>$}410$_{-110}^{+180}$ & \phm{$>$}91$_{-5}^{+6}$ &0.81 $\pm$ 0.04 & 152 $\pm$ 9 & \phm{$>$}840$_{-350}^{+260}$ & \phm{$>$}440$_{-110}^{+160}$\\
J2348-3054 & 457 $\pm$ 49 & $<$65 & $>$150 & \phm{$>$}189$_{-19}^{+22}$ &0.39$_{-0.03}^{+0.04}$ & 238 $\pm$ 25 & $>$380 & $>$290\\
P359-06 & 341 $\pm$ 18 & \phm{$>$}54.0$_{-2.5}^{+2.3}$ & \phm{$>$}126 $\pm$ 8 & \phm{$>$}112$_{-4}^{+5}$ &1.63 $\pm$ 0.07 & 177 $\pm$ 9 & \phm{$>$}316$_{-19}^{+20}$ & \phm{$>$}242 $\pm$ 8\\
\hline
P009-10 & 437 $\pm$ 33 & \phm{$>$}60.7$_{-2.8}^{+2.5}$ & \phm{$>$}125 $\pm$ 12 & \phm{$>$}148$_{-8}^{+9}$ &3.55$_{-0.23}^{+0.25}$ & 227 $\pm$ 17 & \phm{$>$}380 $\pm$ 30 & \phm{$>$}300$_{-14}^{+16}$\\
J0305-3150 & 225 $\pm$ 15 & \phm{$>$}38 $\pm$ 3 & \phm{$>$}86$_{-7}^{+8}$ & \phm{$>$}88.2$_{-2.2}^{+2.4}$ &1.92 $\pm$ 0.07 & 117 $\pm$ 8 & \phm{$>$}272$_{-26}^{+27}$ & \phm{$>$}184 $\pm$ 5\\
P065-26 & 289 $\pm$ 31 & $<$65 & --- & $>$170 &3.3$_{-0.6}^{+0.9}$ & 150 $\pm$ 16 & $>$240 & \phm{$>$}---\\
P167-13 & 519 $\pm$ 25 & \phm{$>$}64.7$_{-1.4}^{+1.3}$ & \phm{$>$}70 $\pm$ 6 & \phm{$>$}145 $\pm$ 3 &1.9 $\pm$ 0.07 & 270 $\pm$ 13 & \phm{$>$}431 $\pm$ 21 & \phm{$>$}277 $\pm$ 6\\
J1306+0356 & 246 $\pm$ 26 & \phm{$>$}53$_{-6}^{+5}$ & \phm{$>$}58 $\pm$ 14 & \phm{$>$}112$_{-6}^{+7}$ &1.64$_{-0.14}^{+0.16}$ & 128 $\pm$ 14 & \phm{$>$}231$_{-28}^{+30}$ & \phm{$>$}215$_{-12}^{+13}$\\
J1319+0950 & 532 $\pm$ 57 & \phm{$>$}39$_{-4}^{+3}$ & \phm{$>$}365$_{-21}^{+28}$ & \phm{$>$}77$_{-4}^{+5}$ &1.73 $\pm$ 0.07 & 280 $\pm$ 30 & \phm{$>$}630 $\pm$ 80 & \phm{$>$}392$_{-20}^{+26}$\\
J1342+0928 & 353 $\pm$ 27 & \phm{$>$}59 $\pm$ 3 & $<$11 & \phm{$>$}99$_{-6}^{+7}$ &2.46$_{-0.14}^{+0.15}$ & 184 $\pm$ 14 & \phm{$>$}308$_{-25}^{+26}$ & \phm{$>$}183$_{-11}^{+13}$\\
P308-21 & 541 $\pm$ 32 & \phm{$>$}77.0$_{-1.4}^{+1.0}$ & \phm{$>$}440$_{-40}^{+30}$ & \phm{$>$}173$_{-11}^{+12}$ &6.3 $\pm$ 0.5 & 281 $\pm$ 17 & \phm{$>$}416 $\pm$ 25 & \phm{$>$}540$_{-33}^{+27}$\\
J2318-3113 & 344 $\pm$ 34 & \phm{$>$}63$_{-6}^{+5}$ & \phm{$>$}110$_{-40}^{+70}$ & \phm{$>$}195$_{-29}^{+62}$ &3.6$_{-0.4}^{+0.5}$ & 179 $\pm$ 18 & \phm{$>$}290 $\pm$ 30 & \phm{$>$}370$_{-50}^{+110}$\\
\enddata
\tablecomments{(1) Name of quasar, (2) FWHM of the \CII\ emission line
  (3-6) inclination, rotational velocity, velocity dispersion and
  effective radius of quasar host as determined from the thin disk
  kinematic modelling. (7-9) circular velocities derived using three
  methods described in Section \ref{sec:rotvel}. $(a)$ Galaxy shows no velocity gradient in its \CII\
  velocity field, and is therefore assumed to be dispersion-dominated.}
\end{deluxetable*}

The primary source of uncertainty in estimating dynamical masses is the uncertainties arising from estimating the circular velocity of the galaxy. Several approaches have been used in the literature for estimating the circular velocity of the gas from the \CII\ emission line. For observations that either do not resolve or only marginally resolve the \CII\ line (i.e., the emission is resolved over less than twice the size of the beam), typically the width of the emission line (i.e., the FWHM) is taken as a proxy for the the circular velocity \citep[e.g.,][]{Walter2004, Wang2013, Decarli2018}. If the velocities are dominated by non-ordered rotation, the circular velocity can be approximated by \citep{Decarli2018}:
\begin{equation} 
\label{eq:vcircdd}
v_{\rm circ} = \sqrt{\frac{3}{16\ln2}}\text{FWHM} \approx 0.52\,\text{FWHM}.
\end{equation}
Whereas if the gas arises from ordered motion, the circular velocity can be approximated by \citep{Ho2007, Wang2013}:
\begin{equation}
\label{eq:vcircrd} 
v_{\rm circ} = 0.75\,\text{FWHM} / \sin i.
\end{equation}
Here $i$ is the inclination of the galaxy and the factor of 0.75 comes from the conversion of the FWHM to the full width at 20\% of the peak intensity, $W_{20}$, for a Gaussian profile \citep[i.e.,][argues that $W_{20}$ is a better tracer of the circular velocity, $v_{\rm circ} = 0.5 W_{20}$, and for a Gaussian profile $W_{20}$/FWHM $\approx 1.5$]{Ho2007}.  Often the inclination cannot be accurately measured, and is taken to be 55$\degr$ \citep{Wang2013, Willott2015, Decarli2018}.

With recent higher resolution observations that resolve the \CII\ emission from $z \gtrsim 6$ quasar host galaxies \citep{Shao2017, Banados2019, Decarli2019, Neeleman2019, Venemans2019, Wang2019}, it has become feasible to provide a better constraint on the morphology and kinematics of the \CII\ emission. This includes a better estimation of the inclination of the galaxy, as well as providing a direct, spatially extended measure of the rotational velocity of the galaxy, either from a position-velocity diagram or direct modeling of the \CII\ emission \citep{Neeleman2019, Pensabene2020}. This rotational velocity is either directly taken as the circular velocity estimate, or the velocity dispersion measurement is taken into account by adding it in quadrature to the circular velocity estimate with some constant of proportionality\footnote{The kinematic estimator, $S_K$, defined as $S_K = \sqrt{K v_{\rm rot}^2 + \sigma_v^2}$ where $K$ is taken between 0.3 - 0.5  \citep{Weiner2006, Kassin2007}, is related to this circular velocity by a linear scaling factor.}, $\eta$:
\begin{equation}
v_{\rm circ}(\eta) = \sqrt{v_{\rm rot}^2 + \eta ~ \sigma_v^2}
\label{eq:vcirceta}
\end{equation}
The value of $\eta$ depends on the mass distribution and kinematics of the galaxy, and can vary between approximately 2 to larger than 6 \citep{Epinat2009, Burkert2010, Turner2017}. Higher values indicate a larger contribution of the velocity dispersion to the dynamical mass estimate, which is often attributed to turbulence within the the ISM of the galaxy. Following \citet{Burkert2010}, if we model the mass distribution as an exponential, turbulent, pressure-supported disk, then $\eta(r) = 2 r / R_{\rm D}$, where $R_{\rm D}$ is the scale length of the exponential disk. At the effective (or half-light) radius of the \CII\ emission, $r = R_{\rm e}$ (see Section \ref{sec:extent}), which for an exponential disk occurs at $r = 1.678R_{\rm D}$, resulting in $\eta = 3.4$. We will take this value as our fiducial value to facilitate comparison with literature values at lower redshift \citep[e.g.][]{Newman2013, Turner2017, FoersterSchreiber2018}. We note, however, that the reported systematic uncertainties in the dynamical mass include a 40 \% contribution that encompasses the expected range in $\eta$.

In Table \ref{tab:vcirc} we list the circular velocity estimates for the three methods discussed. The second column lists the FWHM of the total \CII\ emission as reported in \citet{Venemans2020}. This FWHM measurement is used in column six and seven to estimate the circular velocity, assuming non-ordered and ordered rotation, respectively. This shows that the assumption of non-ordered rotation will yield the lowest estimate to the circular velocity, which on average is a factor of two smaller compared to the assumption of ordered rotation. For the ordered rotation circular velocity estimate, we take the inclination as derived from the kinematic modeling. This inclination is on average $33 \pm 5 \degr$ for those quasars in the undisturbed subsample. This average inclination is smaller than previous estimates \citep{Wang2013, Willott2015}, which is likely due to the exclusion of interacting systems that have larger, but incorrect, inclination estimates. Column eight lists the circular velocity estimate using the results from the kinematic modeling and Equation \ref{eq:vcirceta}. This circular velocity estimate in most cases is bracketed by the other two methods, and will be used to calculate the dynamical masses of the quasar host galaxies.

\subsubsection{Radial Extent for Dynamical Mass Estimate}
\label{sec:extent}

As shown in Equation \ref{eq:mdyn}, the dynamical mass estimate is linearly proportional to the assumed radius of the galaxy. To estimate the radius, previous work have used a variety of different definitions with the most common being either the maximum extent of the \CII\ emission or the radius determined from a 2D Gaussian fit \citep{Wang2013,Willott2015,Decarli2018}. Both these definitions are sub-optimal, as the former is affected by both the S/N and resolution of the observations, whereas the latter underestimates the extent of the emission, because a substantial amount of flux can be observed outside this radius. In this work we assume that the \CII\ emission obeys an exponential radial profile (Section \ref{sec:thindiskmodel}), and the kinematic fitting gives a scale length of the emission profile corrected for the effect of the beam, $R_{\rm D}$. The effective radius, $R_{\rm e}$ (also known as the half-light radius) can be calculated from $R_{\rm D}$ as $R_{\rm e} = 1.678R_{\rm D}$. Lower redshift studies often use $2R_{\rm e} $ as the radius to calculate the dynamical mass \citep[e.g.,][]{Turner2017, FoersterSchreiber2018}, which for an exponential profile contains 85\% of the total flux. For ease of comparison, we will use the same definition in this manuscript. Table \ref{tab:vcirc} contains the estimates for $R_{\rm e}$. The mean effective radius for the undisturbed sample is $1.11 \pm 0.11$ kpc.

\begin{deluxetable*}{lllll}
\colnumbers
\tabletypesize{\small}
\tablecaption{Dynamical, gas, and black hole mass estimates
\label{tab:mdyn}}
\tablehead{
\colhead{Name} &
\colhead{$M_{\rm dyn}$} &
\colhead{${M_{\rm gas, cont}}^a$} &
\colhead{${M_{\rm gas, [CII]}}^b$} &
\colhead{$M_\bullet$} \\
\colhead{} &
\colhead{($10^{10} M_\odot$)} &
\colhead{($10^{10} M_\odot$)} &
\colhead{($10^{10} M_\odot$)} &
\colhead{($10^{9} M_\odot$)}
}
\startdata
P007+04 & $>$1.7($_{-1.2}^{+1.0}$) & 2.2$_{-  0.9}^{+ 11.5}$ & 4.9$_{-  2.4}^{+  4.6}$ & 1.5$_{-  0.5}^{+  2.3}$\\
J0100+2802 & \phm{$>$}10.2$_{-  1.4}^{+  1.6}$($_{-  7.1}^{+  6.1}$) & 1.4$_{-  0.6}^{+  7.6}$ & 12$_{-6}^{+11}$ & 9.73$_{-  0.24}^{+  0.26}$\\
J0109-3047 & \phm{$>$}1.89$_{-  0.27}^{+  0.24}$($_{-  1.3}^{+  1.1}$) & 0.6$_{-  0.25}^{+  3.45}$ & 5.8$_{-  2.8}^{+  5.4}$ & 1.1 $\pm$   0.4\\
J0129-0035 & \phm{$>$}5.4$_{-  2.4}^{+  3.3}$($_{-  3.8}^{+  3.2}$) & 2.3$_{-  1.0}^{+ 11.9}$ & 6.0$_{-  2.9}^{+  5.6}$ & 0.17$_{-  0.09}^{+  0.17}$$^d$\\
J025-33 & \phm{$>$}3.67 $\pm$   0.17($_{-  2.6}^{+  2.2}$) & 2.5$_{-  1.1}^{+ 13.9}$ & 18$_{-9}^{+16}$ & 2.2$_{-  1.1}^{+  2.1}$$^d$\\
P036+03 & \phm{$>$}2.9$_{-  0.7}^{+  1.1}$($_{-  2.1}^{+  1.8}$) & 2.8$_{-  1.1}^{+ 15.4}$ & 10$_{-5}^{+10}$ & 3.7$_{-  0.5}^{+  0.6}$\\
J0842+1218 & \phm{$>$}2.8 $\pm$   0.8($_{-  2.0}^{+  1.7}$) & 0.65$_{-  0.27}^{+  3.45}$ & 2.3$_{-  1.2}^{+  2.2}$ & 1.64$_{-  0.12}^{+  0.15}$\\
J1044-0125 & \phm{$>$}10$_{-3}^{+8}$($_{-  7.3}^{+  6.2}$) & 2.6$_{-  1.1}^{+ 13.7}$ & 5.1$_{-  2.5}^{+  4.8}$ & 3.4 $\pm$   0.5$^c$\\
J1048-0109 & \phm{$>$}4.0$_{-  1.2}^{+  2.2}$($_{-  2.8}^{+  2.4}$) & 3.1$_{-  1.3}^{+ 17.7}$ & 7$_{-3}^{+6}$ & 2.3 $\pm$   0.6\\
J1120+0641 & $>$3.8($_{-2.7}^{+2.3}$) & 0.8$_{-  0.3}^{+  4.8}$ & 3.8$_{-  1.8}^{+  3.5}$ & 2.4$_{-  0.06}^{+  0.05}$$^c$\\
P183+05 & $>$13($_{-9.1}^{+7.8}$) & 5.0$_{-  2.1}^{+ 27.8}$ & 22$_{-11}^{+21}$ & 3.0 $\pm$   0.4\\
P231-20 & \phm{$>$}1.4 $\pm$   0.09($_{-  1.0}^{+  0.8}$) & 4.8$_{-  2.0}^{+ 26.9}$ & 11$_{-5}^{+10}$ & 4.1$_{-  0.7}^{+  1.0}$\\
J2054-0005 & $>$2.9($_{-2.0}^{+1.7}$) & 3.0$_{-  1.2}^{+ 15.8}$ & 10$_{-5}^{+9}$ & 1.48$_{-  0.17}^{+  0.18}$\\
J2100-1715 & \phm{$>$}7.2$_{-  1.9}^{+  2.7}$($_{-  5.0}^{+  4.3}$) & 0.53$_{-  0.22}^{+  2.85}$ & 4.1$_{-  2.0}^{+  3.8}$ & 4.5$_{-  1.5}^{+  0.9}$\\
P323+12 & \phm{$>$}5.3$_{-  0.8}^{+  0.9}$($_{-  3.7}^{+  3.2}$) & 0.25$_{-  0.10}^{+  1.41}$ & 4.5$_{-  2.3}^{+  4.2}$ & 1.12 $\pm$   0.14\\
J2318-3029 & \phm{$>$}7$_{-4}^{+5}$($_{-  5.1}^{+  4.4}$) & 3.0$_{-  1.3}^{+ 16.2}$ & 7$_{-3}^{+6}$ & 1.46$_{-  0.13}^{+  0.15}$\\
J2348-3054 & $>$1.2($_{-0.84}^{+0.72}$) & 2.7$_{-  1.1}^{+ 15.8}$ & 5.5$_{-  2.7}^{+  5.2}$ & 2.4$_{-  0.7}^{+  0.9}$\\
P359-06 & \phm{$>$}4.4 $\pm$   0.4($_{-  3.1}^{+  2.7}$) & 0.8$_{-  0.3}^{+  4.2}$ & 8$_{-4}^{+8}$ & 2.4$_{-  0.4}^{+  0.6}$\\
\hline
P009-10 & \phm{$>$}14.9$_{-  1.7}^{+  1.9}$($_{- 10.4}^{+  8.9}$) & 3.4$_{-  1.4}^{+ 18.1}$ & 28$_{-14}^{+26}$ & 2.4$_{-  0.5}^{+  0.6}$$^c$\\
J0305-3150 & \phm{$>$}3.03$_{-  0.19}^{+  0.21}$($_{-  2.1}^{+  1.8}$) & 5.9$_{-  2.4}^{+ 33.2}$ & 18$_{-9}^{+17}$ & 0.54$_{-  0.12}^{+  0.11}$\\
P065-26 & \phm{$>$}--- & 1.3$_{-  0.6}^{+  7.2}$ & 5.3$_{-  2.6}^{+  5.0}$ & 4.6$_{-  0.5}^{+  0.6}$\\
P167-13 & \phm{$>$}6.8 $\pm$   0.4($_{-  4.8}^{+  4.1}$) & 1.0$_{-  0.4}^{+  5.3}$ & 17$_{-8}^{+16}$ & 0.3$_{-  0.12}^{+  0.08}$\\
J1306+0356 & \phm{$>$}3.5$_{-  0.5}^{+  0.6}$($_{-  2.5}^{+  2.1}$) & 0.7$_{-  0.29}^{+  3.69}$ & 3.5$_{-  1.7}^{+  3.3}$ & 2.1$_{-  0.06}^{+  0.07}$\\
J1319+0950 & \phm{$>$}12.4$_{-  1.3}^{+  1.7}$($_{-  8.7}^{+  7.4}$) & 5.0$_{-  2.1}^{+ 26.6}$ & 12$_{-6}^{+12}$ & 1.89 $\pm$   0.11\\
J1342+0928 & \phm{$>$}3.8$_{-  0.5}^{+  0.6}$($_{-  2.7}^{+  2.3}$) & 0.48$_{-  0.19}^{+  3.04}$ & 4.1$_{-  2.0}^{+  3.8}$ & 0.91$_{-  0.13}^{+  0.14}$\\
P308-21 & \phm{$>$}86$_{-13}^{+11}$($_{- 59.9}^{+ 51.3}$) & 1.2$_{-  0.5}^{+  6.4}$ & 10$_{-5}^{+10}$ & 1.69$_{-  0.35}^{+  0.20}$\\
J2318-3113 & \phm{$>$}23$_{-7}^{+14}$($_{- 16.4}^{+ 14.1}$) & 0.38$_{-  0.16}^{+  2.09}$ & 4.9$_{-  2.4}^{+  4.6}$ & 0.53$_{-  0.26}^{+  0.52}$$^d$\\
\enddata
\tablecomments{(1) Name of quasar, (2) dynamical mass estimate, (3) molecular gas mass estimate from the continuum flux, (4) molecular gas mass estimate from the \CII\ emission, and (5) black hole mass measured from \ion{Mg}{2}, except where noted. $(a)$ Mass determined from converting the continuum flux density into a dust mass and then assuming a constant dust-to-gas ratio. $(b)$ Mass determined from converting the \CII\ luminosity directly into a molecular mass using the conversion of \citet{Zanella2018}. $(c)$ BH mass determined from the \ion{C}{4} line. $(d)$ BH mass determined assuming Eddington luminosity accretion.}
\end{deluxetable*}

There are two caveats to using the above radius as the extent of the emission. First, we note that \citet{Novak2020} find that the average \CII\ emission profile of $z \gtrsim 6$ quasar host galaxies is actually best described by a double exponential profile with a bright, compact component and a faint, extended component. The faint, extended component contains between 10-20\% of the total flux and thus our single exponential estimate for $R_{\rm e}$ could underestimate the true effective radius by approximately 40\%. We add this systematic uncertainty in $R_{\rm e}$ to the systematic uncertainties in the dynamical mass estimate for each galaxy.

The second caveat is that $R_{\rm e}$ in this work is calculated for \CII\ emission, while for lower redshift observations the effective radii is derived from rest-frame UV/optical observations. These tracers need not be equal, because \CII\ traces gas with a range of physical conditions \citep[e.g.,][]{Pineda2013}, whereas UV/optical observations trace the stellar light. As mentioned previously, quasar host galaxies remain difficult to observe in the rest-frame UV/optical with current facilities, making it difficult to compare spatial extents of \CII\ and the rest-frame UV/optical emission. For other high redshift galaxies where this comparison is possible, the results vary from approximately similar extent \citep{Neeleman2020} to \CII\ being roughly twice as extended \citep[e.g.,][]{Matthee2019}. It is possible the latter studies recover the faint, extended component mentioned in the previous caveat. Nevertheless, we will conservatively add a 50\% systematic uncertainty to the lower bound of the dynamical mass estimate to account for this possible overestimation of the effective radius by using \CII\ as a tracer.

\subsubsection{Dynamical Mass Estimates}
\label{sec:mdyn}

With the above assumptions, we can now estimate the dynamical mass of $z \gtrsim 6$ quasar host galaxies at $2R_{\rm e}$ using the circular velocity estimate from Equation \ref{eq:vcirceta}. The dynamical masses are given in Table \ref{tab:mdyn}, where the first set of uncertainties are the propagated uncertainties on the derived parameters, and the second set of uncertainties in parentheses are the systematic uncertainties as described in the previous sections. For those galaxies that show an undisturbed morphology, the dynamical masses range between 1.2 and $>$13 $\times 10^{10} M_\odot$, with a mean mass of $5.0 \pm 0.8\,(\pm 3.5) \times 10^{10} M_\odot$.

\section{Discussion}
\label{sec:discussion}

\subsection{Large Velocity Dispersions}
\label{sec:highdisp}

\begin{figure*}
\includegraphics[width=\textwidth]{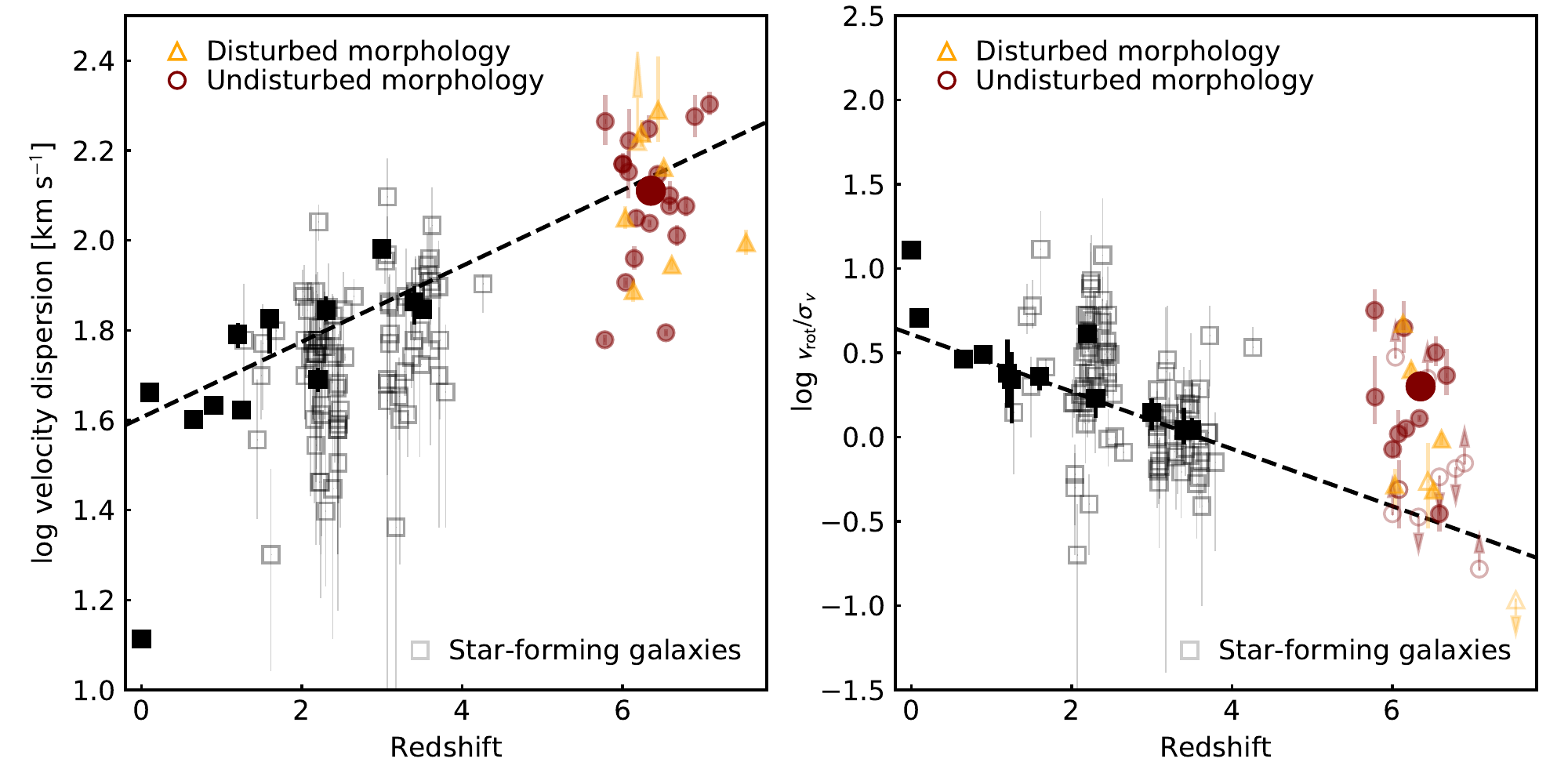}
\caption{\textit{left:} Redshift evolution of the velocity dispersion. The $z \gtrsim 6$ quasar host galaxy sample is shown in colored symbols and is divided into two subsamples based on the \CII\ morphology. The large solid circle is the average velocity dispersion for the undisturbed \CII\ subsample. The squares are low redshift measurements for massive, near-main sequence galaxies for both full sample averages (solid black squares) as well as individual measurements (open gray squares). An exponential extrapolation based on the low-redshift data is shown by the dashed line, indicating that the $z \gtrsim 6$ quasar sample falls on this simple extrapolation. \textit{right:} Redshift evolution for the ratio between rotation velocity and velocity dispersion, \vsigrat. Here the open symbols in the $z \sim 6$ quasar sample mark upper and lower limits. The $z \gtrsim 6$ quasar host galaxies have \vsigrat\ comparable to $z \sim 2-3$ galaxies, which is higher than a simple extrapolation based on the lower redshift data (dashed line).
\label{fig:vrotsigv}}
\end{figure*}

A remarkable feature of the \CII\ emission lines from $z \gtrsim 6$ quasar host galaxies is their intrinsically large velocity dispersion. The mean velocity dispersion for the undisturbed sample is $129 \pm 10$ \kms. The velocity dispersion of the sample, which has been corrected for beam smearing effects (Section \ref{sec:models}), is plotted as a function of redshift in Figure \ref{fig:vrotsigv}. Also plotted in Figure \ref{fig:vrotsigv} are the velocity dispersion measurements for both individual galaxies \citep[e.g.,][]{Livermore2015,Turner2017,FoersterSchreiber2018} and sample averages \citep[for references, see compilation in][]{Turner2017} at lower redshifts. This shows that $z \sim 6$ quasar host galaxies have a velocity dispersion that is, on average, larger than any other lower redshift galaxy sample. 

At $z \lesssim 4$, previous works have found that the velocity dispersion evolves with redshift. Extrapolating an exponential fit to this evolution out to $z \gtrsim 6$ shows that the quasar population falls on this extrapolation (Figure \ref{fig:vrotsigv}). This suggests that although the velocity dispersions are higher than any previously measured samples, this could simply be due to the higher redshift of the velocity dispersion measurements. However, we caution that interpreting this extrapolation is challenging, in part because of the different sample selection criteria. Most low redshift samples consist of star-forming galaxies that fall near or on the main sequence, whereas the quasar sample by definition have a strong AGN at its center. If the extrapolation is valid, this would imply that the AGN does not significantly alter the velocity dispersion measurement of the quasar host galaxy.

The large velocity dispersion also affects the shape of the \CII\ emission profile for quasars. As has been noted before \citep[e.g.,][]{Decarli2018, Neeleman2019}, the profile of the \CII\ emission from $z \gtrsim 6$ quasar host galaxies can be approximated with a Gaussian function. This need not be the case, as emission from a gaseous disk often results in a `double-horned' emission profile \citep[e.g.,][]{DeBlok2014}. The lack of any `double-horned' profiles within the sample is caused by the large velocity dispersion, as the separation between potential peaks need to be greater than approximately twice the velocity dispersion of the peaks (i.e., $v_{\rm rot} \sin i \gtrsim \sigma_v$). In addition, the exponential profile of the \CII\ emission could cause some of the emission being sampled preferentially from the rising part of the rotation curve \citep{DeBlok2014}. Together these effects could explain the ubiquity (but not necessity) of Gaussian profiles in emission profiles of $z \gtrsim 6$ quasar host galaxies.

We end this section with the observation that the subsample of 8 galaxies that show no sign of a velocity gradient have a significantly larger velocity dispersion (mean velocity dispersion of $153 \pm 32$ \kms) than the 10 galaxies that show a velocity gradient (mean velocity of $95 \pm 25$ \kms). Besides the difference in velocity structure, no other obvious differences are found between these two subsamples. In particular, they have similar inclination angles, spatial extents and black hole masses. This suggests that the inherent velocity structure in these two subsamples of quasar host galaxies is different. One possible scenario that is qualitatively consistent with these observations is that roughly half (8 out of 18) of the $z \gtrsim 6$ quasar host galaxies have \CII-emitting gas that is dominated by turbulent motions, which prevents the gas from settling into a disk. The turbulence could be a temporary or transitionary state caused by energy injection from either a recent merger or strong AGN feedback. This temporary state scenario is consistent with the small proximity zones for at least two of the eight galaxies within the subsample \citep[J0100$+$2802 and J2100$-$1715;][]{Eilers2017, Eilers2020}, since short proximity zones are an indication that the quasar phase of the SMBH has just started. However, further observations of the proximity zones for all quasars in the sample are needed to explore if the potential turbulence seen in these host galaxies is a temporary state related to the recent start of the quasar.

\subsection{Dispersion- and Rotation-Dominated Galaxies}
\label{sec:vrotvsig}

\begin{figure*}[t]
\includegraphics[width=\textwidth]{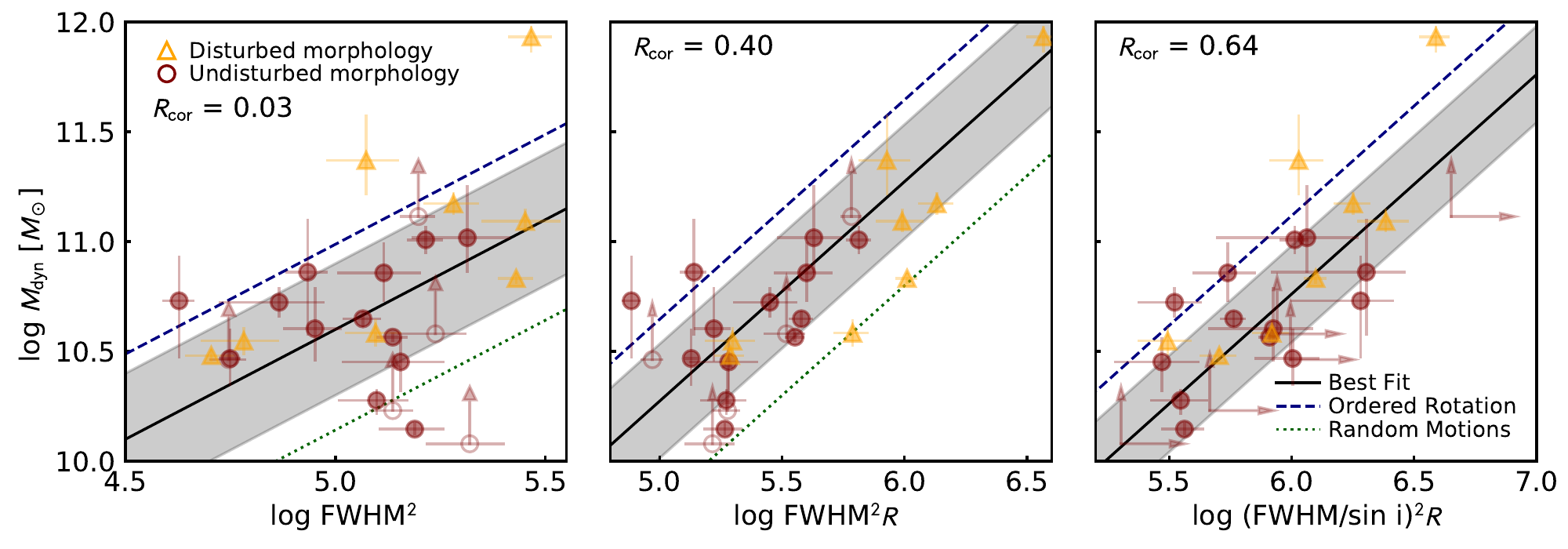}
\caption{Dynamical mass estimates determined from modeling the \CII\ emission line plotted against several observables available in low resolution observations. In the left panel, we plot the dynamical mass against the square of the FWHM of the \CII\ emission line (FWHM$^2$). There is no strong correlation found between these two quantities with a correlation coefficient, $R_{\rm cor}$, of 0.03. A best fit line is shown by the solid black line, where the uncertainties on the line (gray shaded region) encompasses  $>$68\,\% of the data. Also shown are the dynamical mass estimates assuming the mass is dispersion dominated (dotted line; Equation \ref{eq:vcircdd}) and rotationally supported (dashed line; Equation \ref{eq:vcircrd}), which bracket the dynamical mass estimates. The middle panel shows the dynamical mass against the observable, FWHM$^2 R$, where $R$ is the radial extent of the emission. There is a better correlation between these quantities, resulting in a better estimate of the dynamical mass. Finally the best estimate for the dynamical mass can be obtained from the observable (FWHM/sin i)$^2 R$ (right panel). Equations \ref{eq:mdynf2} -- \ref{eq:mdynf2ri} give the conversions from these observables to the dynamical mass.
\label{fig:mdynfwhm}}
\end{figure*}

A common empirical diagnostic used to determine the rotational support of a galaxy is the ratio between the rotational velocity and the velocity dispersion \citep[\vsigrat; e.g.,][]{Epinat2009, Burkert2010, Turner2017}, where higher \vsigrat\ correspond to galaxies whose kinematics are more dominated by rotational motions. Varying cutoffs have been used to separate galaxy samples into `rotation-dominated' systems and `dispersion-dominated' systems, the most popular being either \vsigrat\ $> 1$ \citep{Epinat2009, Newman2013, Turner2017}, and \vsigrat\ $> 3$ \citep{Burkert2010, FoersterSchreiber2018}. The former cut-off corresponds to the visual classification used to divide the sample into those galaxies that show a velocity gradient and those that lack a velocity gradient (see Section \ref{sec:veldispfield}), with 8 out of 18 galaxies having \vsigrat\ $< 1$. We consider these systems to be dispersion-dominated. The remaining 10 galaxies have \vsigrat\ $> 1$, with four galaxies (J0129$-$0035, P036$+$03, J2054$-$0005, and J2318$-$3029) having \vsigrat\ $> 3$. These latter galaxies have the lowest velocity dispersions of the full sample, as well as smooth velocity gradients (Fig. \ref{fig:VFieldSingle}), and satisfy all criteria for being `velocity-dominated' disk galaxies \citep[see][]{Wisnioski2015}.

We can compare \vsigrat\ for the $z \gtrsim 6$ quasar host galaxies with similar measurements for the lower redshift samples described in Section \ref{sec:highdisp}. For these samples,  \vsigrat\ is shown as a function of redshift in Fig. \ref{fig:vrotsigv}. We find that the sample of $z \gtrsim 6$ quasar host galaxies have \vsigrat\ consistent with the samples at $z \sim 2 - 3$. This is unlike the velocity dispersion measurement, which is much higher among $z \gtrsim 6$ quasar host galaxies. Indeed, when we exponentially extrapolate the low redshift evolution in \vsigrat, the $z \gtrsim 6$ quasar host galaxies fall above this extrapolation. This immediately implies that most of the sample has a higher rotational velocity than expected from the extrapolation. This is not surprising, as the $z \gtrsim 6$ quasar host galaxies are thought to be biased toward the most massive, evolved galaxies at this redshift.

\subsection{Estimating Dynamical Mass from Unresolved Observations}
\label{sec:mdynest}

We have used resolved observations to provide some of the most accurate measurements of the dynamical mass of $z \gtrsim 6$ quasar host galaxies to date. In this section, we explore how these dynamical mass estimates compare to mass estimates gleaned from low spatial resolution observations for which only an integrated spectrum can be obtained. We perform this comparison for three different cases depending on the available observational data. 

The first case assumes only an integrated FWHM measurement is available. In this case, we assume that the galaxy has a radial extent of 2.2~kpc and an inclination of 33$\degr$, which are the mean extent and inclination of the full sample. Under these assumptions, we can convert the FWHM measurement into a dynamical mass estimate using Equations \ref{eq:vcircdd} and \ref{eq:vcircrd}. These dynamical mass estimates are displayed with the dashed and dotted lines in the left-most panel of Figure \ref{fig:mdynfwhm}. Also shown in this panel are the dynamical mass estimates for the full sample. This panel shows that the dynamical mass estimates from Equations \ref{eq:vcircdd} and \ref{eq:vcircrd} roughly bracket the dynamical mass estimates from the kinematic modeling. We obtain a best-fit line to the data set of
\begin{equation}
\label{eq:mdynf2}
\left(\frac{M_{\rm dyn}}{M_\odot}\right) = (4.0^{+4.0}_{-2.0} (^{+2.4}_{-2.8})) \times 10^5~\left(\frac{\text{FWHM}}{\text{km s}^{-1}}\right)^2.
\end{equation}
Here the first set of uncertainties on the constant of proportionality defines the range that encompasses roughly 68 \% of the data (gray-shaded region in Figure \ref{fig:mdynfwhm}), and the second set of uncertainties in parentheses are the systematic uncertainties on the dynamical mass estimate (see Section \ref{sec:dynmass}). 

The second case assumes we have a measurement of the integrated FWHM of the galaxy and the radial extent, but not a measurement of the inclination, which we assume to be 33$\degr$. In this case, the dynamical mass estimates using Equations \ref{eq:vcircdd} and \ref{eq:vcircrd} again bracket the dynamical mass estimates, and the best fit line to the data is:
\begin{equation}
\label{eq:mdynf2r}
\left(\frac{M_{\rm dyn}}{M_\odot}\right) = (1.9^{+1.5}_{-0.8} (^{+1.1}_{-1.3})) \times 10^5~\left(\frac{\text{FWHM}}{\text{km s}^{-1}}\right)^2\left(\frac{R}{\text{kpc}}\right).
\end{equation}
Numerically, this equation is very similar to the theoretical equation under the assumption that the emission arises from gas inside a virialized system \citep[see][]{Bothwell2013, DessaugesZavadsky2020}.

Finally, the last case assumes we have a measurement of the integrated FWHM, radial extent and a measurement of the inclination. In this case, Equation \ref{eq:vcircrd} again predicts, on average, dynamical masses that are too high compared to the measured values. We find a best fit regression line of:
\begin{equation}
\label{eq:mdynf2ri}
\left(\frac{M_{\rm dyn}}{M_\odot}\right) = (5.8^{+3.8}_{-2.3} (^{+3.5}_{-4.0})) \times 10^4~\left(\frac{\text{FWHM}/\sin i}{\text{km s}^{-1}}\right)^2\left(\frac{R}{\text{kpc}}\right).
\end{equation}
We note that this last equation corresponds to an empirical circular velocity estimate of $v_{\rm circ} = (0.50^{+14}_{-11})\,\text{FWHM} / \sin i$. This empirical estimate for the circular velocity is lower compared to the physically motivated correlation of Equation \ref{eq:vcircrd}. This is presumably driven by those systems that are dominated by random motions.

Equations \ref{eq:mdynf2} -- \ref{eq:mdynf2ri} provide empirical prescriptions with which dynamical masses of $z \gtrsim 6$ quasar host galaxies can be estimated, and where the dynamical mass estimates become more accurate when more observational data is available. As shown in the left-most panel of Fig. \ref{fig:mdynfwhm}, there appears to be little correlation between the FWHM and the dynamical mass estimates. This correlation increases with increasing observational data (i.e, the addition of the radial extent and the inclination), where the spread of the data around the best fit correlation decreases by approximately 50\,\% allowing for more accurate dynamical mass estimates. We note that the systematic uncertainties start to dominate the dynamical mass uncertainties when accurate measurements of FWHM, spatial extent and inclination are available.

\subsection{Comparison between Dynamical and SMBH Mass}
\label{sec:mdynmbh}

\begin{figure}
\includegraphics[width=0.48\textwidth]{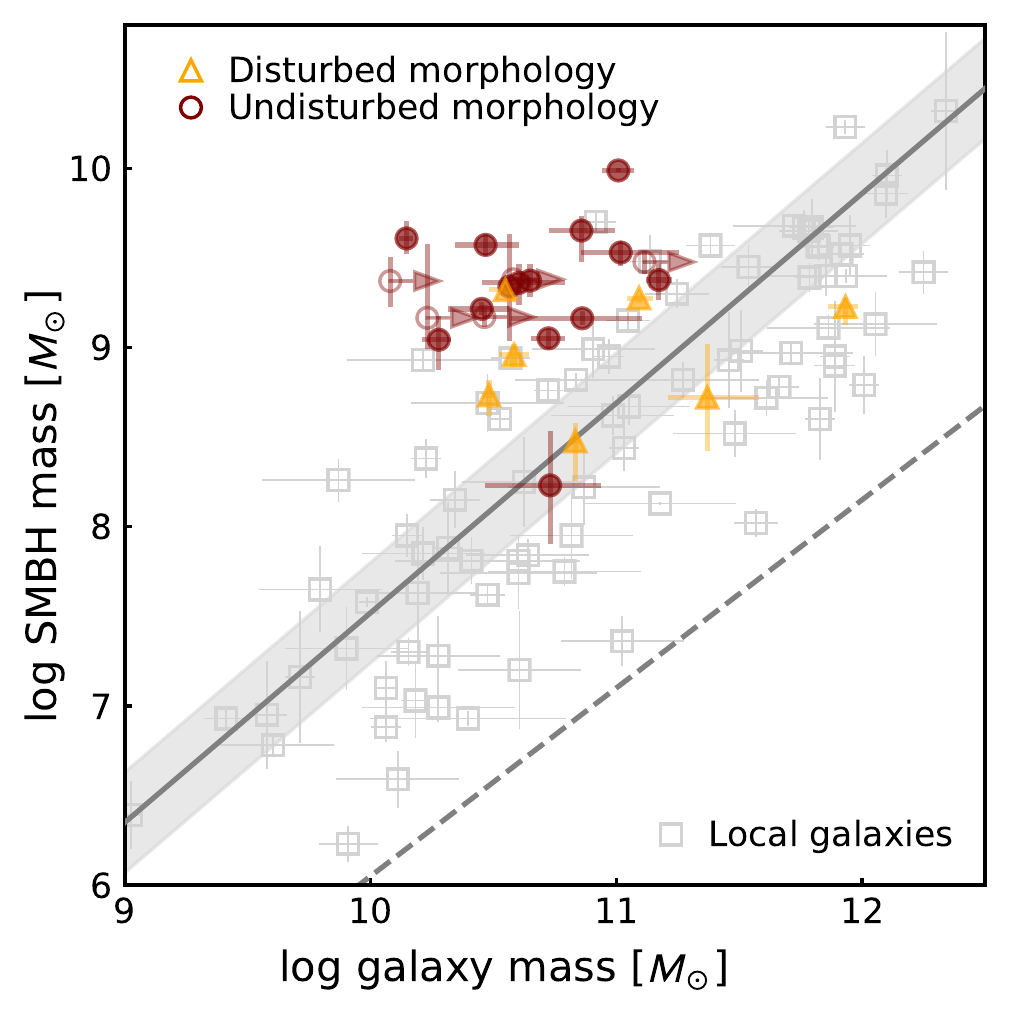}
\caption{Black hole mass versus the dynamical mass estimate for the $z \gtrsim 6$ quasar sample. Also shown is the black hole mass versus bulge mass for local galaxies \citep{Kormendy2013, DeNicola2019} on the same axes. The fit to this data from \citet{Kormendy2013} is shown as a solid line where the shaded region marks the 1$\sigma$ scatter seen in the data around this correlation. Also shown is the fit to the total stellar mass versus black hole mass for local AGNs \citep[dashed line;][]{Reines2015}. The quasar sample is apportioned into two groups based on the \CII\ morphology, and are generally observed above the locally derived correlation.
\label{fig:mdyn}}
\end{figure}

One of the goals of this manuscript is to compare the dynamical mass estimate of quasar host galaxies with the mass of the SMBH at their center. This comparison is shown in Figure \ref{fig:mdyn}. Nearly all of the SMBH mass estimates are taken from a recent study analyzing NIR spectra from a large sample of $z \gtrsim 6$ quasars \citep[][Farina et al. in prep.]{Schindler2020}, except for P167$-$13 \citep{Mazzucchelli2017} and J1342$+$0928 \citep{Onoue2020}. For 21 of the quasars in our sample, the \ion{Mg}{2} emission line was used for the SMBH mass determination. However, for six quasars the \ion{Mg}{2} emission lines could not be used. For three quasars the SMBH mass was estimated from the \ion{C}{4} line, whereas for the remaining three quasars the black hole mass estimate was determined from the luminosity assuming that the SMBH accretes at the Eddington limit \citep[e.g.,][]{DeRosa2011}. The black hole masses and tracer used are given in Table \ref{tab:mdyn}. The sample has a mean SMBH mass of $2.2 \pm 0.5$ $\times 10^{9} M_\odot$.

Because of the limited range in SMBH masses (21 quasars have a SMBH mass between $1 - 5 \times 10^9 M_\odot$), we do not attempt to search for a correlation between SMBH mass and dynamical mass within the sample. Instead, we compare these measurements with the relationship between bulge mass and black hole mass for local galaxies \citep[e.g.,][]{Kormendy2013, DeNicola2019}. If we assume that the dynamical mass and the bulge mass are roughly equivalent between these studies, then we find that the sample of $z \gtrsim 6$ quasars occupy a different region of parameter space compared to local galaxies. On average, we find that for comparable black hole masses, the host galaxies of $z \gtrsim 6$ quasars have smaller dynamical masses compared to their low redshift counterparts. This is not a new conclusion, previous work with either lower resolution data \citep{Walter2003, Ho2007, Willott2015, Venemans2016, Decarli2018} or smaller sample sizes \citep{Shao2017, Izumi2018, WangF2019, Wang2019, Pensabene2020} as well as studies of the stellar light from lower redshift galaxies \citep[e.g.,][]{Decarli2010} have made similar claims. This study expands on these results by providing a much larger sample of high resolution observations (at least twice as large as any previous study). Comparing the mean of the $z \gtrsim 6$ quasar sample, in which the uncertainties are dominated by the systematics described in the previous sections, with the local relationship of \citet{Kormendy2013} shows that the dynamical masses of these quasars are smaller by about a factor of seven, which is a 2.4$\sigma$ deviation from the local correlation. We note that this result remains unchanged if we only consider those quasars that have a black hole estimate from \ion{Mg}{2}.

There are two critical assumptions in the above analysis that need to be explored. First is the assumption that the dynamical masses and bulge masses are comparable measurements, and therefore plotting them on a common axis in Figure \ref{fig:mdyn} is sensible. Naively we would expect that the dynamical mass estimate of a galaxy is strictly greater than the stellar mass of the bulge as the former takes into account all mass components (i.e., stars, gas, dark matter, etc.). The local correlations between stellar mass and SMBH mass should therefore shift to the right when converted to dynamical mass. This would make the discrepancy between local galaxies and the $z \gtrsim 6$ quasar sample even more significant \citep[see e.g.,][]{Reines2015}. However, this naive assumption ignores systematic uncertainties in the derivation of the stellar mass estimates and dynamical mass estimates. As an example, the local galaxy NGC 3091 \citep{Rusli2013} has a stellar mass estimate that is about 0.4 dex higher than the dynamical mass estimate using the estimators described in this paper. It was exactly this systematic uncertainty that prompted \citet{Kormendy2013} to ignore any mass determinations based on kinematic arguments, and only use bulge mass estimates based on mass-to-light ratios. A detailed discussion on the magnitude of this systematic is beyond the scope of this work, but if the results from NGC 3091 are typical, then much of the tension between high redshift quasars and the local bulge mass - black hole mass relation could be alleviated by accounting for this systematic.

\begin{figure*}
\includegraphics[width=\textwidth]{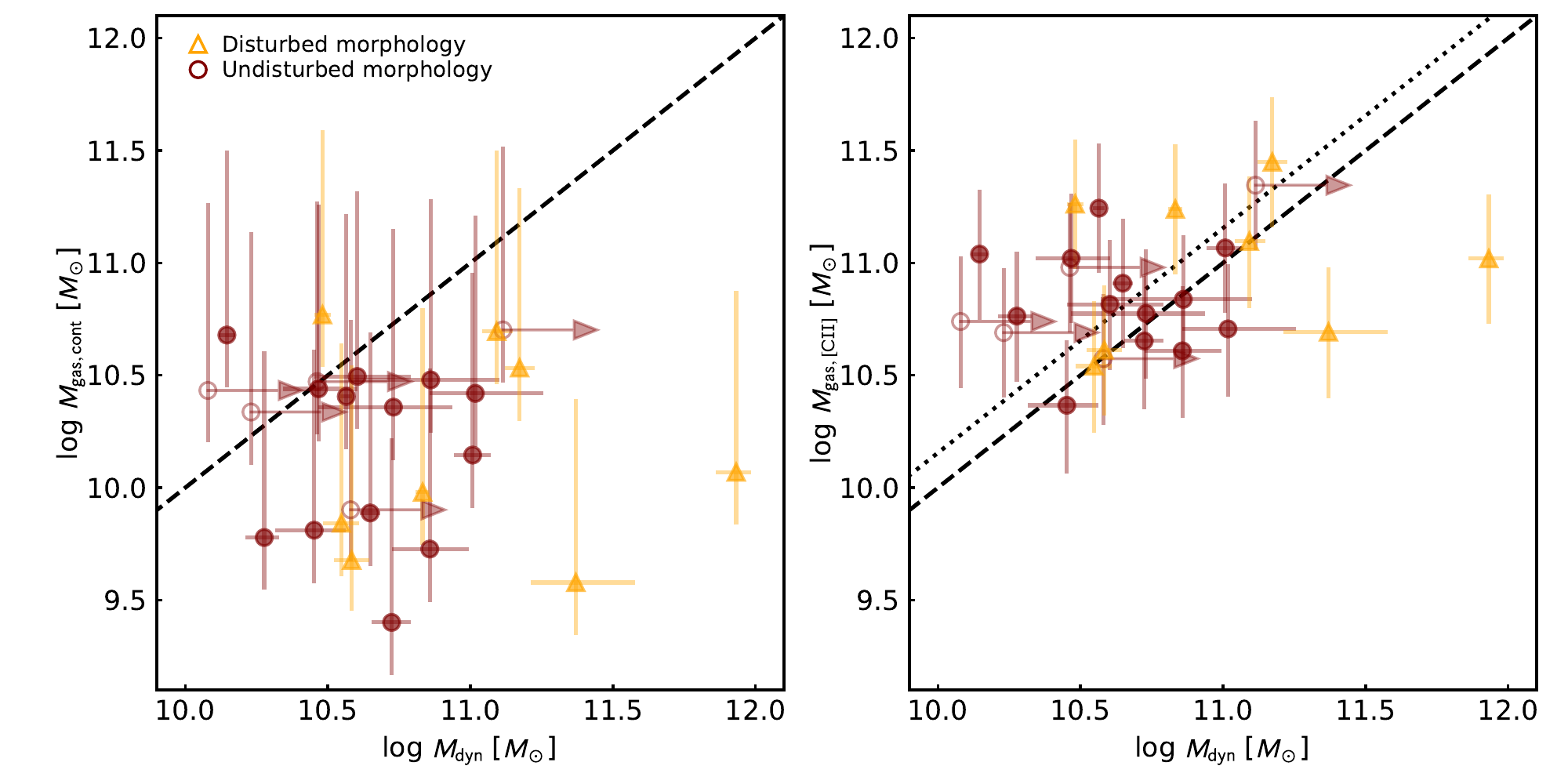}
\caption{Molecular mass estimates compared to the dynamical mass either from the dust continuum emission \citep[left panel: assuming standard conversion estimates e.g.,][]{Venemans2016}, or from the scaling between \CII\ luminosity and molecular mass \citep[right panel:][]{Zanella2018}. In the latter case, the \CII-based molecular gas mass estimate is unphysically higher than the dynamical mass estimate, suggesting that the scaling relationship derived in \citet{Zanella2018} for main-sequence galaxies at $z \sim 2$ does not apply to $z \gtrsim 6$ quasar host galaxies. We find that the gas mass estimates derived from the dust continuum are lower than the dynamical mass estimates, as expected, because the dynamical mass accounts for all mass components within the given radius, including the gas.
\label{fig:mcomp}}
\end{figure*}

The second assumption is that the quasar sample used in this study is representative of the massive black hole ($>10^9 M_\odot$) population at $z \gtrsim 6$. In particular, that the sample is not biased toward lower-mass host galaxies. In local galaxies, massive black holes ($>10^9 M_\odot$) are found in massive (10$^{12} M_\odot$) galaxies. We assert here that we are not biased against observing these massive systems neither because of dust obscuration nor due to difficulty detecting \CII\ emission from these galaxies. If a large fraction of bright quasars were obscured at high redshift, we would expect to see a steepening of the bright end slope of the quasar luminosity function as more quasars would remain undetected. No such steepening is observed \citep{Kulkarni2019, Schindler2019}. We also do not expect that \CII\ emission would be harder to observe from more massive quasar host galaxies, because low resolution \CII\ observations of $z \gtrsim 6$ quasars achieve a success rate of approximately 85\% \citep{Decarli2018}, and none of these observed quasars show much extended emission \citep[see][]{Novak2020}. 

An alternative view of the second assumption is that the highly luminous quasars could probe the most massive end of the black hole distribution for the dynamical mass range probed. This view is corroborated by the results of \citet{Izumi2019}, who use marginally resolved \CII\ observations to find that less UV-luminous $z \sim 6$ quasars fall closer or even below the local bulge mass - black hole mass relation. However, this does not explain the lack of massive ($\approx$10$^{12} M_\odot$) quasar host galaxies in our black hole mass-based sample. Only if such massive galaxies do not occur at $z \sim 6$, can we reconcile these observations with this assumption.

\subsection{Gas Masses versus Dynamical Masses}
\label{sec:MgasMdyn}

With accurate dynamical masses for a sample of $z \gtrsim 6$ quasar host galaxies, we can start exploring how much the gas contributes to the total mass of these systems. Estimates for the molecular mass are determined in two ways. For the first method, we convert the dust continuum flux measurement into a dust mass assuming the method described in detail in \citet{Venemans2016} accounting for the effects of the CMB \citep{DaCunha2013}. In particular we assume a dust temperature of T$_{\rm d}$ = 47 K, an emissivity index of $\beta$ = 1.6, and a dust mass opacity coefficient of $\kappa_\lambda = 0.77 (850 \mu{\rm m}/\lambda)^\beta$~cm$^2$~g$^{-1}$ \citep{Dunne2000}. Dust masses are converted into molecular masses assuming a dust-to-gas ratio of 100 and a molecular-to-total gas mass fraction of 0.75. Uncertainties are calculated assuming T$_{\rm d}$ ranges between 30 and 60~K and $\beta$ ranges between 1.4 and 1.8. 
For the second method, we convert the \CII\ luminosity into a molecular mass estimate assuming the conversion given in \citet{Zanella2018} with $\alpha_{\rm [CII]} = 31~L_\odot/M_\odot$ as calibrated for $z \sim 2$ main-sequence galaxies. Uncertainties are dominated by uncertainty in the conversion factor, which we take to be 0.3 dex \citep[as suggested in][]{Zanella2018}. The molecular mass estimates for both methods are given in Table \ref{tab:mdyn}. 

In Figure \ref{fig:mcomp} we compare these molecular mass estimates to the dynamical mass estimates derived in this manuscript. We find that if we apply the molecular mass conversion from the \CII\ luminosity assuming the conversion in \citet{Zanella2018}, we get molecular mass estimates that are on-average greater than the dynamical mass estimates with a median molecular-to-dynamical mass ratio of 1.7. Even if we account for the fact that at most 20\,\% of the emission occurs outside the radius used for the dynamical mass estimate (Section \ref{sec:extent}), and we account for possible systematic underestimation of the dynamical mass, the molecular mass estimates remain above the dynamical mass estimates. This suggests that the conversion factor derived for main-sequence galaxies at $z \sim 2$ is not valid for quasar host galaxies at $z \gtrsim 6$. 

When we use the dust continuum emission to estimate the molecular mass, we find a molecular-to-dynamical mass ratio of 0.3. If we again conservatively assume that 30\,\% of the molecular mass is outside the radius used for the dynamical mass estimate and we account for the uncertainties in both the molecular and dynamical mass estimate, we still find that the molecular gas accounts for $>$10\,\% of the total dynamical mass of the system. In passing, we note that if we use the dust continuum-based molecular mass estimates advocated in \citet{Scoville2016}, we would get molecular mass estimates above the mass estimates based on \CII, much larger than the dynamical mass estimates. Therefore, independent of the molecular mass estimate, the implication is that molecular gas contributes a non-negligible mass to $z \gtrsim 6$ quasar host galaxies, roughly an order of magnitude larger than the mass contribution from the SMBH. 

High molecular gas mass fractions are consistent with the results found at lower redshift in main-sequence star-forming galaxies \citep{DessaugesZavadsky2020, Price2020}, and the empirical observation that gas mass fractions increase with redshift \citep{Carilli2013, Tacconi2020}. High gas fractions are further supported by simulations \citep[e.g.,][]{Lupi2019}, which suggests that the gas mass could be comparable to the stellar mass in these systems. Detection of the stellar mass of quasar host galaxies will have to await high resolution, rest-frame near-infrared imaging from space. However, current observations of the gas-to-dynamical mass fraction are consistent with this picture, if the dark matter content of these galaxies within the inner few kpc is negligible.

\section{Summary and Concluding Remarks}

In this manuscript, we explore the kinematics of a sample of 27 $z \gtrsim 6$ quasar host galaxies observed in \CII\ emission at a resolution of approximately $0\farcs25$. We have used our publicly available python-based fitting code, \texttt{qubefit}, to fit the kinematics of the \CII\ emission line. The main results from this analysis are summarized below.

\begin{itemize}
\item One third (9 out of 27) of quasar host galaxies have disturbed \CII\ emission profiles. This includes quasars that show distinct emission components, but have emission that encompasses the individual components, and systems with complicated velocity profiles (See Fig. \ref{fig:VFieldComplex}). About one third of quasar host galaxies (10 out of 27) show a smooth velocity gradient consistent with the emission arising from a rotating disk, and the remaining one third (8 out of 27) show a velocity profile without a clear velocity gradient, which is consistent with the emission arising from a dispersion-dominated system. It is interesting to note that a similar division is also seen among resolved observations of $z \sim 2$ massive, main-sequence galaxies \citep{Burkert2010}.
 
\item We see no evidence for a deviation from a constant velocity dispersion and constant rotational velocity across the extent of the \CII\ emission. This implies that with the resolution of our observations, we cannot accurately determine the rising part of the velocity curve of these systems, nor do we find evidence for any elevated dispersion at the center of the galaxy due to the gravitational influence of the SMBH.

\item The mean velocity dispersion of the sample is $129 \pm 10$~\kms\ with a median of 123~\kms. Although much larger than previous lower redshift measurements \citep{Turner2017, FoersterSchreiber2018, Price2020}, we assert this is due to the higher redshift of the observations, as the velocity dispersion measurements agree with a simple exponential extrapolation with redshift based on the lower redshift data. These observations support the hypothesis that galaxies at higher redshift are more turbulent and show less rotational support than galaxies at low redshift. If the extrapolation is correct, it would obviate the need for any contribution to the velocity dispersion from the SMBH. The subsample of dispersion-dominated galaxies show, on average, larger velocity dispersions. This suggests that these quasar host galaxies have gas that is turbulent, which prevents the gas from settling into a disk. The source of this turbulence needs to be examined further, but it could be temporary and caused by energy injection from a recent merger or AGN feedback.

\item We measure the dynamical masses for all galaxies in the sample, although we caution that the estimates for galaxies in the disturbed sample are not very reliable due to non-gravitationally supported motions. We find a mean dynamical mass for the undisturbed subsample of $5.0 \pm 0.8~(\pm 3.5) \times 10^{10} M_\odot$, where the uncertainty in parentheses is the systematic uncertainty. We also provide empirical formulae for estimating the dynamical mass from unresolved (or barely resolved) observations (Equations \ref{eq:mdynf2} -- \ref{eq:mdynf2ri}). 

\item Comparing the dynamical and SMBH mass, we find that $z \gtrsim 6$ quasar host galaxies fall above the locally derived correlation \citep{Kormendy2013}. This indicates that $z \gtrsim 6$ quasar host galaxies are less massive than implied from the mass of the SMBH. This result is robust even if we account for the large systematic uncertainty in the estimation of the dynamical mass. One possible caveat to this result is the implicit assumption that the dynamical mass is a good proxy for the bulge mass in these systems. This assumption has large systematic uncertainties that could assuage the discrepancy between the local galaxies and the high redshift sample.

\item Comparing the dynamical and molecular mass indicates that the gas mass fractions of $z \gtrsim 6$ quasar host galaxies is high, with a conservative lower limit to the molecular gas mass contribution of $>$10\,\%. These high molecular gas mass fractions are consistent with the observational trends seen in lower redshift galaxies. Even if we assume that the dark matter mass is negligible within the inner few kpc of the galaxy, and the remainder of the mass is all within stars, then molecular gas would contribute a significant fraction of the total baryonic mass of the galaxy.

\end{itemize}

These observations reveal that the kinematics of $z \gtrsim 6$ quasar host galaxies vary. Although about a third of all quasar host galaxies appear to have \CII-emitting gas that is constrained to a disk, there are equally as many sources that show the gas being disturbed by a recent merger and those that have turbulent gas dynamics sufficient to suppress the gas from forming a disk. This suggests that there is not a single precipitating event that causes the active quasar phase of $z \gtrsim 6$ SMBHs. Clearly this statement is very speculative with the current observations. Higher resolution observations of the \CII\ emission line using ALMA, and detailed rest-frame near-infrared observations using the James Webb Space Telescope are needed to confirm its validity.  

\acknowledgements
This paper makes use of the following ALMA data: ADS/JAO.ALMA $\#$2012.1.00240.S, $\#$2012.1.00882.S, $\#$2013.1.00273.S, $\#$2015.1.00339.S, $\#$2015.1.00692.S, $\#$2016.A.00018.S, $\#$2016.1.00544.S, $\#$2017.1.00396.S, $\#$2017.1.01301.S, and $\#$2018.1.00908.S. ALMA is a partnership of ESO (representing its member states), NSF (USA) and NINS (Japan), together with NRC (Canada), NSC and ASIAA (Taiwan), and KASI (Republic of Korea), in cooperation with the Republic of Chile. The Joint ALMA Observatory is operated by ESO, AUI/NRAO and NAOJ. Ma.N., Ml.N., B.P.V and F.W. acknowledge support from ERC Advanced grant 740246 (Cosmic\texttt{\char`_}Gas).

\facilities{ALMA}

\software{Qubefit (this work), Emcee \citep{ForemanMackey2013}, Astropy \citep{Astropy2013, Astropy2018}, SciPy \citep{Virtanen2020}, NumPy \citep{Harris2020}, CASA \citep{McMullin2007}, Matplotlib \citep{Hunter2007}, Corner \citep{ForemanMackey2016}}.

\bibliography{Bib.bib}

\appendix
\section{Channel maps}
\label{sec:channelmaps}
\begin{figure*}[!b]
\centering
\includegraphics[width=0.74\textwidth]{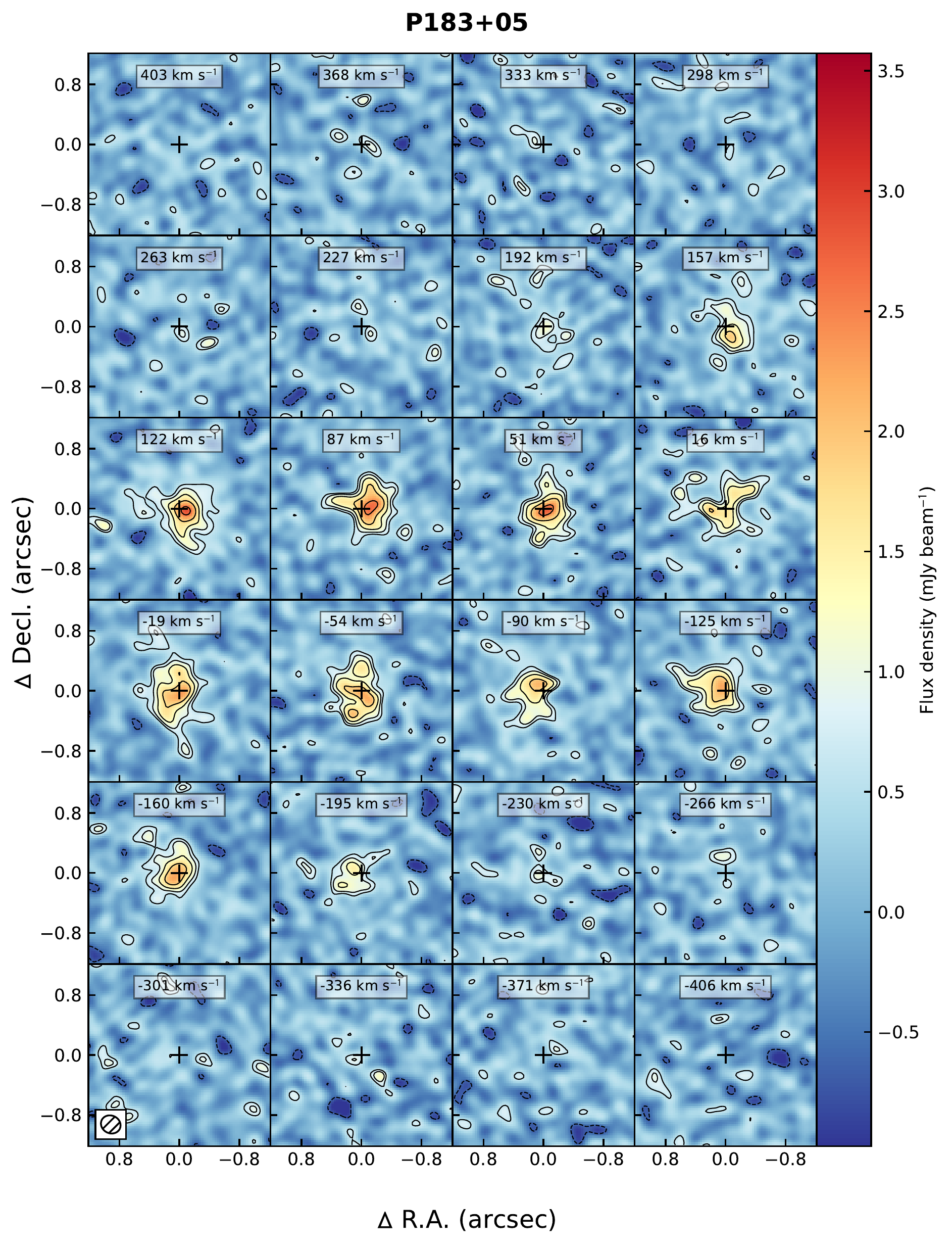}
\caption{Channel maps of the \CII\ emission line for P183$+$05. The image is centered on the optical position of the quasar (black plus sign) and velocities are relative to the redshift of the \CII\ emission line as determined in \citet{Venemans2020}. Contours start at 2$\sigma$ and increase in powers of $\sqrt{2}$ with negative contours at the same level dashed. The ALMA synthesized beam is shown in the bottom left inset. The channel maps for all quasars of the sample (27 images) are available in the online journal or from the corresponding author directly.
\label{fig:chanmap}}
\end{figure*}

In this appendix we show the channel maps of the \CII\ emission line for all quasars in the sample (Figure \ref{fig:chanmap}).

\newpage

\section{Analysis of the Kinematic Modeling}
\label{sec:kinanalysis}

\begin{figure*}[!b]
\includegraphics[width=\textwidth]{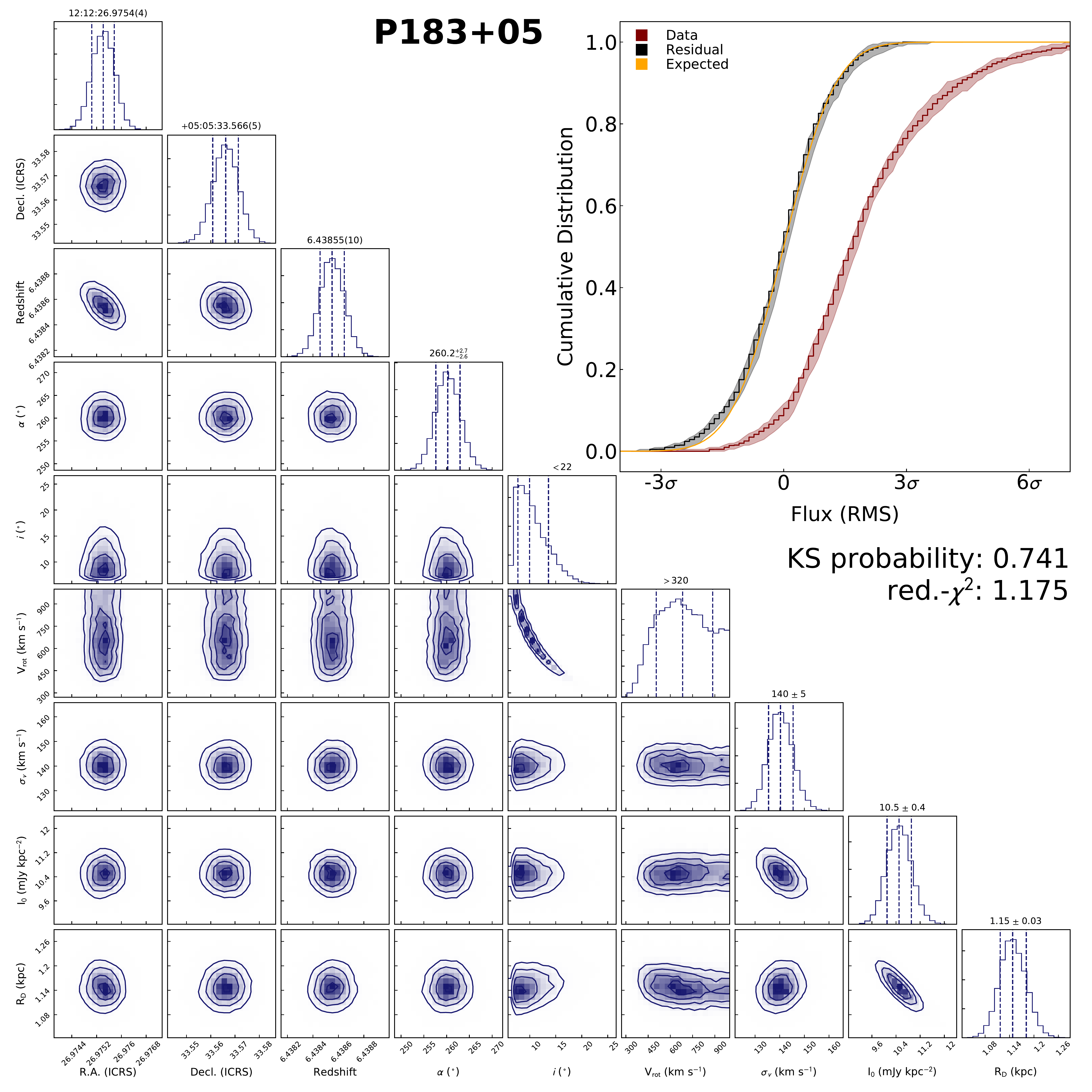}
\caption{Corner plot for the thin disk model of P183$+$05. The bottom panels show the dependencies between the parameters of the model as well as PDFs for each individual parameter. The dashed lines in the panels with the PDFs mark the 16$^{\rm th}$, 50$^{\rm th}$ and 84$^{\rm th}$ percentile of each distribution. The best estimates with uncertainties (or 3$\sigma$ limits) are reported above the PDF of each parameter. The inset displays the cumulative distribution function of the flux within the mask described in Section \ref{sec:gof} for both the data and the residual. Also shown is the expected distribution of the residual, if the noise was purely Gaussian. In this example case, the residual agrees well with the expected Gaussian noise distribution, suggesting a good fit, as also suggested by a low reduced-$\chi^2$ value and high  Kolmogorov-Smirnov probability (shown directly below the inset). The full set of corner plots for both models and all quasars (27 images) are available in the online journal or from the corresponding author directly.
\label{fig:cornerplot}}
\end{figure*}

To illustrate the fitting procedure performed for all of the quasars in the sample, we here describe the process for one such quasar, P183$+$05. This quasar was chosen as an example case because the \CII\ emission is extended, it is detected at good significance, and its observational properties are typical for those quasars that have extended, resolved \CII\ emission. When running the fitting code, the first MCMC-chain was initialized by visually finding initial conditions that produced reasonable reduced $\chi^2$-values (red.-$\chi^2 < 2$). The second chain was initialized with parameter values that were chosen randomly but within 20\,\% of the first chain. After removal of the burn-in period and validation that the two chains converged to the same PDF, the two chains were combined to form a single large chain with a total of 420\,000 realizations. 

We plot the results of the fitting procedure for the thin disk model of P183$+$05 in Figure \ref{fig:cornerplot}. We note that a similar figure from the fitting procedure of the dispersion-dominated bulge model, as well as the results from the fitting procedure of all 27 quasar host galaxies are available in the online journal. The bottom left plots of Figure \ref{fig:cornerplot} contain PDFs of the individual parameters, as well as 2D PDFs that highlight possible correlations between individual parameters \citep[in a so-called corner plot; e.g.,][]{ForemanMackey2013}. The diagonal plots display the PDF of each individual parameter, and are mostly Gaussian. For these parameters we can determine a median value as well as the 1$\sigma$ uncertainties (defined by the values that denote the 16$^{\rm th}$ and 84$^{\rm th}$ percentile of the PDF). Two parameters of the thin disk model of P183$+$05, the inclination ($i$) and the rotational velocity ($v_{\rm rot}$), display non-Gaussian distributions. We note that these parameters are strongly anti-correlated, as is shown in their 2D PDF, where small inclinations can lead to very large rotational velocities. For these two parameters we therefore only report an upper and lower limit. Additional weaker correlations are seen between right ascension and declination as well as the two emission profile parameters, $I_0$ and $R_{\rm D}$. The former correlation is due to the major axis being aligned with the right ascension direction, and a shift in redshift would imply a shift in the center of the emission in order to reproduce the observed velocity field, whereas the latter correlation is due to roughly conserving the total emission from the galaxy.

In the top right inset of Figure \ref{fig:cornerplot}, we show the results from the two goodness-of-fit analyses, i.e., the KS statistic and the reduced $\chi^2$ values (Section \ref{sec:gof}). We also show a cumulative plot of the flux inside the union of the fitting mask and the 1$\sigma$ contour (see Section \ref{sec:gof}). The observational data within this region shows many pixels above 3$\sigma$ (i.e., the line detection). However, after subtracting the model from the data, the residual shows almost no $>$3$\sigma$ features. Indeed if we compare the distribution of the flux values in the residual data cube with the expected Gaussian noise distribution, as calculated from channels without any emission, we see excellent agreement. This agreement is numerically captured by the KS-statistic, and indicates that the noise properties of the residual are consistent with the rest of the cube. This suggest that the model can accurately describe the observational data.

\begin{figure*}[!t]
\centering
\includegraphics[width=0.77\textwidth]{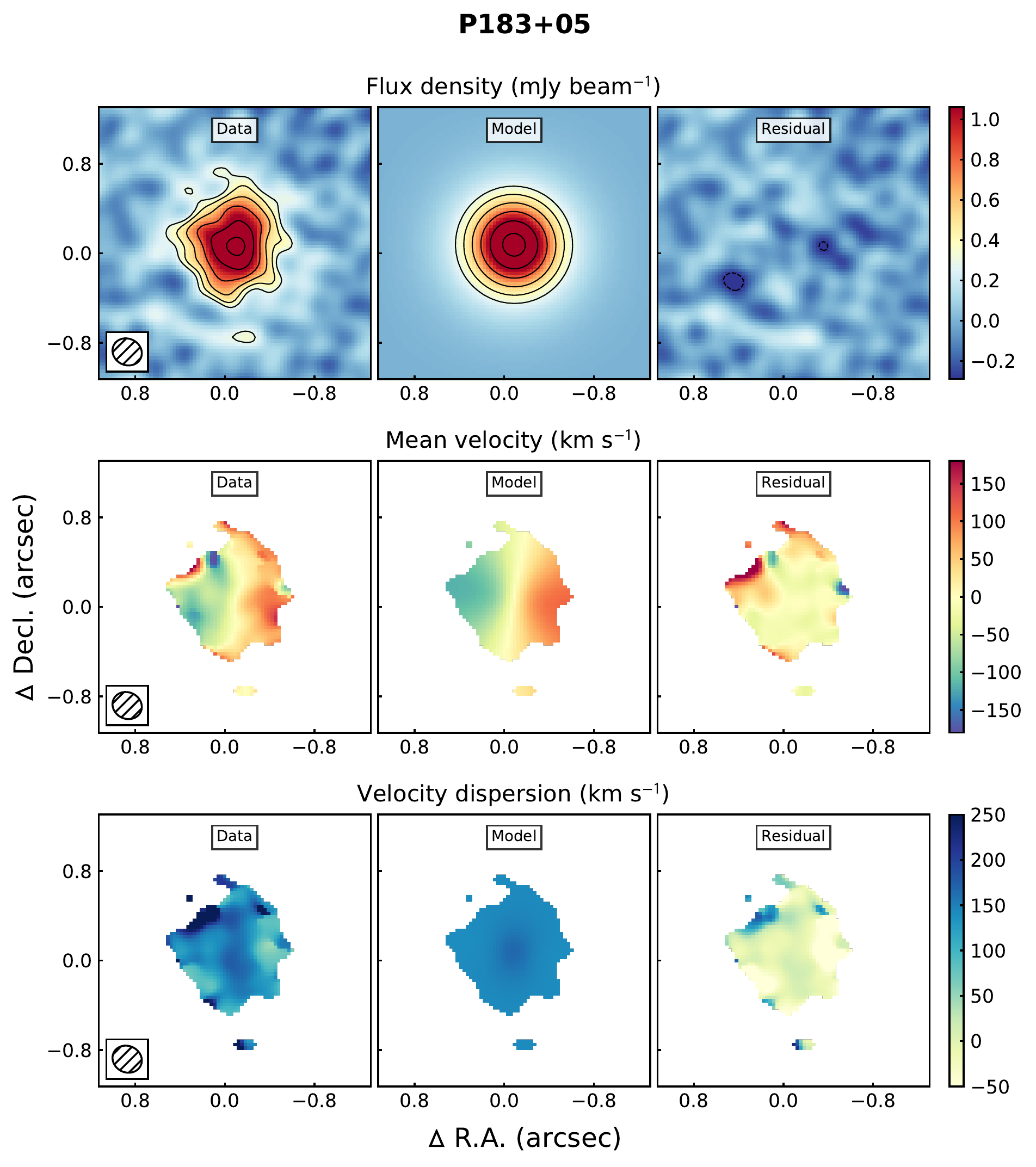}
\caption{Comparison of the flux density, mean velocity and velocity dispersion between the data for P183$+$05 and the best-fit thin disk model. Top row shows the flux density of the \CII\ emission. The middle row shows the velocity field, and the bottom row the velocity dispersion of the galaxy. The resolution of the data is shown by the bottom left inset. All panels are shown on the same scale. Contours in the flux density start at 3$\sigma$ and increase in powers of $\sqrt{2}$. The figures for both models and all quasars (27 images) are available in the online journal or from the corresponding author directly.
\label{fig:dmcomp}}
\end{figure*}

\begin{figure*}[!b]
\centering
\includegraphics[width=0.77\textwidth]{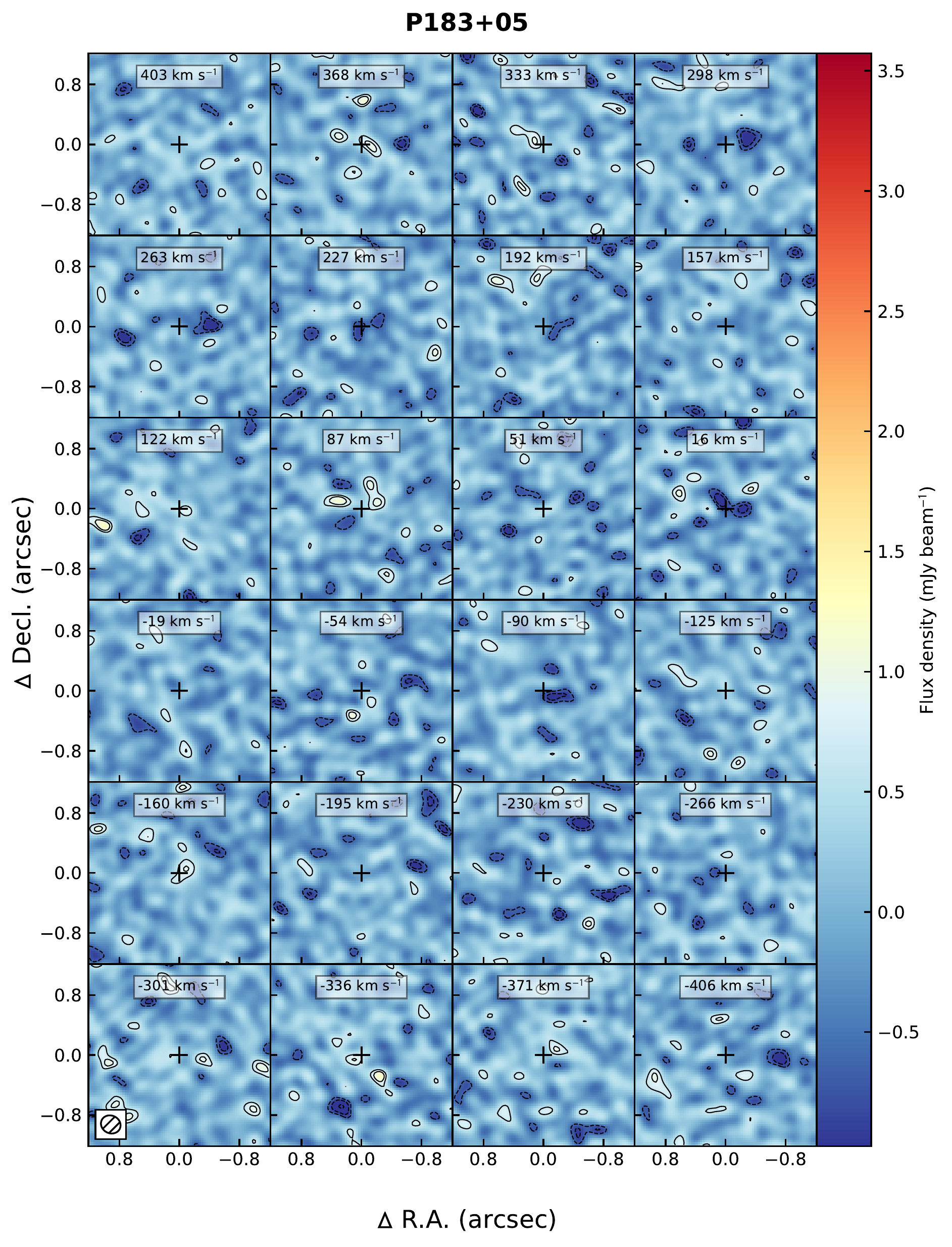}
\caption{Channel maps of the residual of the \CII\ emission line for P183$+$05 after applying the best-fit thin disk model, which has been convolved with the ALMA beam. To facilitate comparison with the channels maps of the data (Figure.~\ref{fig:chanmap}), positions annotations and contour levels have been kept the same. The channel maps of the residual using both the thin disk model and dispersion dominated bulge model for all quasars of the sample (27 images) are available in the online journal or from the corresponding author directly.
\label{fig:cmres}}
\end{figure*}

We provide several other visual diagnostics of the fitting procedure in Figures \ref{fig:dmcomp} and \ref{fig:cmres}. Figure \ref{fig:dmcomp} shows the flux density, mean velocity and velocity dispersion fields for the data, the beam-convolved model, as well as the residuals after subtracting the convolved model from the data. The residual channel maps, which are created after subtracting the convolved model from the data for each channel, are shown in \ref{fig:cmres}.

\section{Two Methods for Deriving Mean Velocity and Velocity Dispersion Fields}
\label{sec:velfieldmethods}

Generating a velocity and velocity dispersion field from a 3D data cube has been discussed in detail in \citet{Deblok2008}. Although many approaches for generating these fields exists, previous studies of high redshift galaxies have typically either estimated the fields from calculating the moments of the data cube along the spectral direction \citep[e.g.,][]{Wang2013, Rybak2019}, or from spectral line fitting of the emission line at each spatial position in the data cube \citep[e.g.,][]{Debreuck2014, Shao2017}. 

\begin{figure*}
\includegraphics[width=\textwidth]{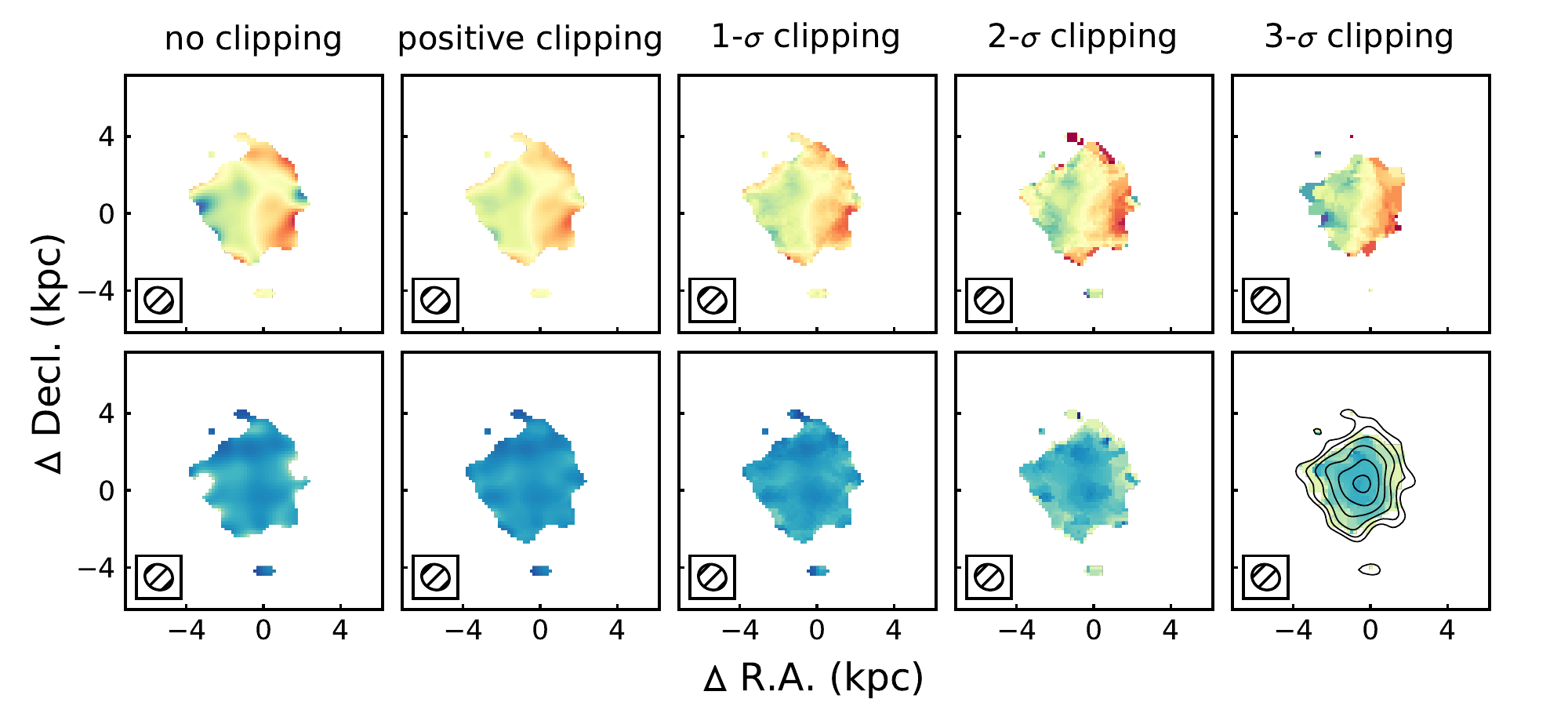}
\caption{Velocity and velocity dispersion fields for the host galaxy of quasar P183$+$05. The inset in each panel shows the ALMA synthesized beam for these observations. All of the fields have been masked to show only pixels with a total \CII\ intensity detected at $>$3$\sigma$. From left to right the threshold used for clipping the data cube is increased from no clipping and clipping all negative values to clipping all data below multiples of the noise RMS. When no clipping is performed, the velocity dispersion field is undefined for certain regions in which the sum of Equation \ref{eq:mom12} is negative. When the data cube is clipped at $>$2$\sigma$, the velocity dispersion field starts to decrease artificially at the edges and start to resemble the intensity field (shown by contours in the bottom rightmost panel). 
\label{fig:MomClip}}
\end{figure*}

The first method is fast and easy to implement. The first (M1) and second (M2) moment along the spectral direction of a data cube in the discreet limit can be calculating using the following formulae:
\begin{equation}
\label{eq:mom12}
\text{M}1 = \frac{\sum^n_i I_iv_i}{\sum^n_i I_i} \hspace{1.5cm} \text{and} \hspace{1.5cm} \text{M}2 = \sqrt{\frac{\sum^n_i I_i(v_i - \text{M}1)^2}{\sum^n_i I_i}}
\end{equation}
Here $v_i$ is the velocity and $I_i$ is the intensity of the $i^{\rm th}$ spatial element (often termed `spaxel') in the data cube. The sum is over all $n$ spaxels along the spectral direction that satisfy the chosen selection criteria. These selection criteria varies among different studies. Most often the spectral region is confined to those frequencies that show some line emission. In addition, often the data cube is clipped below a certain value to remove noise in the data from dominating the velocity field. The choice of this clipping can significantly alter the velocity field. To illustrate this point, we have applied the moment method to the data cube of P183$+$05 using different clipping thresholds from no clipping to clipping all spaxels below a threshold of three times the noise of the observations (Figure \ref{fig:MomClip}). This figure shows that although the overall velocity field remains roughly consistent, velocity measurements of individual spaxels can vary significantly based on the choice of clipping. The choice of clipping results in even larger variations in the velocity dispersion field. When the data gets clipped at $\gtrsim$ 2$\sigma$, the edges of the \CII\ emission appear to have lower velocity dispersion. This is an artifact of clipping. Because the edges contain very few spaxels above the clipping threshold, the width of the line gets reduced. When this happens, the velocity dispersion field will roughy start to resemble the intensity field of the line (bottom rightmost panel Figure \ref{fig:MomClip}). Aggressive clipping can therefore significantly alter the velocity dispersion field, resulting in low S/N regions having artificially low velocity dispersions.

The second method relies on fitting a functional form to the spectrum at each pixel. The advantages and disadvantages for several choices for this functional form are discussed in \citet{Deblok2008}. However, for high redshift observations previous studies have almost exclusively relied on a single Gaussian function \citep[e.g.,][]{Debreuck2014, Shao2017, Venemans2019}. In this manuscript we use the fitting routines in \texttt{astropy} \citep{Astropy2013, Astropy2018} to fit a single Gaussian to the spectrum of each individual spaxel. For initial guesses of the Gaussian function, we take the results from the moment images described above, but we note that the end results are largely independent of the exact choice of the initial guesses. The final routines are made available in \texttt{qubefit}. The main advantage of using a fitting function is that no input needs to be clipped. As a result, this method yields consistent results independent of the S/N of the data. The main drawback with this method is the assumed function form of the spectral line. This functional form might not correspond to the true spectral profile, although for quasars the spectra appear Gaussian (Section \ref{sec:highdisp}), or beam-smearing could distort the shape of the spectral profile. 

\begin{figure*}
\includegraphics[width=\textwidth]{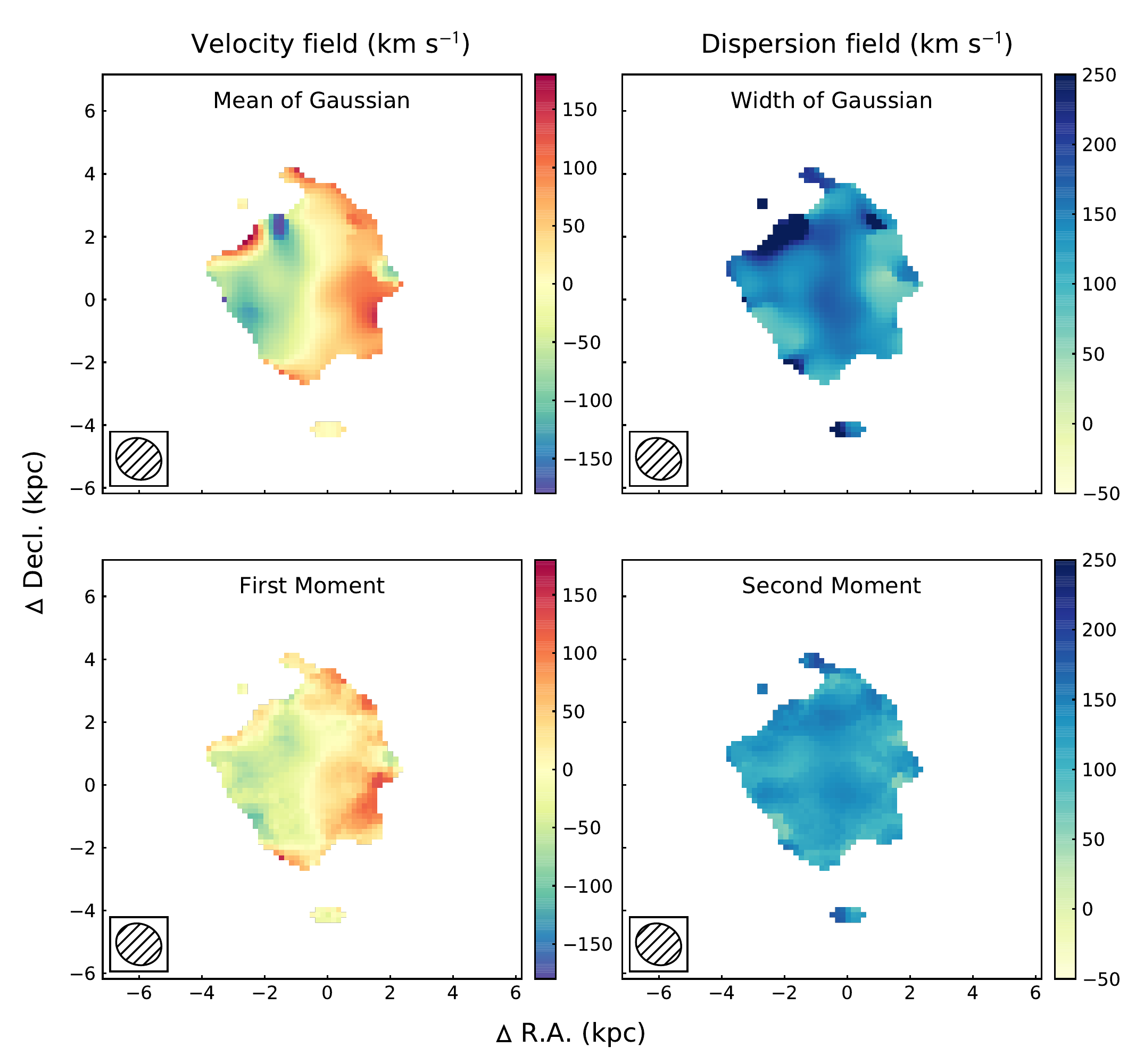}
\caption{Comparison between the two different methods for generating velocity and dispersion fields for the \CII\ emission line from the host galaxy of quasar P183$+$05. The top row shows the results from fitting a single Gaussian function to the spectrum of each individual spaxel. The data is masked to show only pixels that have been detected at $>$3$\sigma$ in the total intensity field. The bottom row is the first moment of the data cube, where the data has been clipped at $>$1$\sigma$. Both methods yield roughly the same velocity and velocity dispersion field. Since the Gaussian fitting routine is independent of the S/N of the data, we opt to use this method in our analysis. The ALMA syntehsized beam for these observations is shown in the bottom left inset. 
\label{fig:MomComp}}
\end{figure*}

We show the velocity and the velocity dispersion field for one quasar (P183$+$05) generated using both methods in Figure \ref{fig:MomComp}. Visually the velocity and velocity dispersion field appear roughly similar, although there are some low S/N regions that have substantial deviations in either field. There is no evidence that the assumption of a Gaussian profile is systematically skewing the measurements in either field. Because of the robust nature of the Gaussian fitting and its independence on the noise properties of the data, we opt to use this method for estimating the velocity and velocity dispersion fields for all quasar host galaxies in our sample.

\end{document}